\documentclass[aps,prd,preprint,superscriptaddress,amsmath,amssymb,showpacs]{revtex4-1}
\usepackage{dcolumn}
\usepackage{graphicx}
\usepackage{mathrsfs}
\usepackage{float}
\usepackage{physics}
\usepackage{footnote}
\usepackage[colorlinks=true,allcolors=blue]{hyperref}
\usepackage{xcolor}
\colorlet{myblue}{blue}

\newcommand{\myDelta}{\Delta\hspace{-2.3ex}\raisebox{0.3ex}{\Large$\sim$}}
\newcommand{\mydelta}{\Delta\hspace{-1.78ex}\raisebox{0.5ex}{\normalsize$\rule{0.32cm}{1pt}$}}

\begin{document}
\title{Attractors and Late Time Asymptotics in a Generalized Relativistic Second Order Spin Hydrodynamics}
\author{Qi Zhou}
\email{qizhou@m.scnu.edu.cn}
\affiliation{State Key Laboratory of Nuclear Physics and Technology, Institute of Quantum Matter, South China Normal University, Guangzhou 510006, China}
\affiliation{Guangdong Basic Research Center of Excellence for Structure and Fundamental Interactions of Matter, Guangdong Provincial Key Laboratory of Nuclear Science, Guangzhou 510006, China}

\author{Duan She}
\email{sheduan@hnas.ac.cn}
\affiliation{Institute of Physics, Henan Academy of Sciences, Zhengzhou 450046, China}
\affiliation{Key Laboratory of Quark \& Lepton Physics (MOE) and Institute of Particle Physics,Central China Normal University, Wuhan 430079, China}

\author{Ben-Wei Zhang}
\email{bwzhang@mail.ccnu.edu.cn}
\affiliation{Key Laboratory of Quark \& Lepton Physics (MOE) and Institute of Particle Physics,Central China Normal University, Wuhan 430079, China}

\author{Enke Wang}
\email{wangek@scnu.edu.cn}
\affiliation{State Key Laboratory of Nuclear Physics and Technology, Institute of Quantum Matter, South China Normal University, Guangzhou 510006, China}
\affiliation{Guangdong Basic Research Center of Excellence for Structure and Fundamental Interactions of Matter, Guangdong Provincial Key Laboratory of Nuclear Science, Guangzhou 510006, China}

\date{\today}

\begin{abstract}
We investigate the attractor of spin density in relativistic spin hydrodynamics using Zubarev’s non-equilibrium statistical operator formalism in the spin probe limit.  
We derive the (0+1)D Bjorken flow equations and the associated attractor equation while retaining second order gradient corrections in the relevant relaxation constitutive equations including couplings associated with nonlinear response and nonlocal memory effects.
We analyze the early time fixed point structure and analytically determine the early time attractor solution, thereby clarifying branch selection and the role of different dynamical corrections.
We find that source-like driving terms modify the leading correction to the attractor solution without changing the fixed point structure, whereas self feedback terms involving the rotational stress tensor modify the dominant balance and modify the early time fixed point structure.
We further analyze the late time asymptotics in the conformal limit and show that the newly added terms modify the algebraic prefactor and logarithmic phase shift, while leaving the leading late time decay unchanged.
These results provide a unified picture of early time attractor and late time asymptotics in the conformal limit.
\end{abstract}

\maketitle

\section{Introduction}

In non-central relativistic heavy ion collisions, the large orbital angular momentum generated in the reaction zone can induce strong vortical structures in the quark gluon plasma (QGP). Through spin and orbit coupling, such vortical motion may polarize quarks and gluons along the direction of the system's angular momentum, thereby giving rise to global spin polarization~\cite{Liang:2004ph,Liang:2004xn,Becattini:2013fla}. 
This picture has been supported by a series of experimental measurements, including the first observation of global $\Lambda$ hyperon polarization by the STAR Collaboration in Au+Au collisions at RHIC~\cite{STAR:2017ckg}, subsequent measurements of multistrange hyperons such as $\Xi^-$ and $\Omega^-$~\cite{STAR:2020xbm}, and the observation of spin alignment for vector mesons including $\phi$ and $K^{*0}$~\cite{STAR:2022fan}. 
These developments have established spin phenomena as an important frontier in high energy nuclear physics. Nevertheless, several issues remain unresolved, such as the difference between $\Lambda$ and $\bar{\Lambda}$ polarization~\cite{STAR:2023nvo,Peng:2022cya}, the rapidity dependence of global polarization~\cite{Liang:2019pst}, the origin of hyperon local polarization~\cite{Karpenko:2016jyx}, and the flavor dependence of vector meson spin alignment~\cite{ALICE:2022dyy}, see reviews \cite{Gao:2020vbh,Becattini:2020ngo,Becattini:2022zvf,Hidaka:2022dmn,Becattini:2024uha} and references therein.

Relativistic hydrodynamics provides an effective framework for describing the collective evolution of the hot and dense QGP created in heavy ion collisions. Motivated by experimental evidence for spin polarization, relativistic spin hydrodynamics extends this framework by incorporating spin degrees of freedom and angular momentum dynamics. In this theory, energy momentum conservation and total angular momentum conservation must be treated consistently, leading to additional evolution equations for the spin tensor. Various approaches have been developed to formulate spin hydrodynamics, 
including  kinetic theory based approaches~\cite{Florkowski:2017ruc,Florkowski:2018fap,Li:2019qkf,Hongo:2022izs,Bhadury:2020puc,Shi:2020htn,Bhadury:2020cop,Peng:2021ago,Weickgenannt:2022zxs,Weickgenannt:2022qvh,Weickgenannt:2022jes},
entropy current analysis~\cite{Hattori:2019lfp,Fukushima:2020ucl,Li:2020eon,She:2021lhe,Hongo:2021ona,Cao:2022aku,Hu:2022azy,Biswas:2023qsw},
holographic duality~\cite{Gallegos:2020otk,Garbiso:2020puw},
effective hydrodynamic theories~\cite{Gallegos:2021bzp,Gallegos:2022jow,Montenegro:2017rbu,Montenegro:2020paq}, and Zubarev's formalism of non-equilibrium statistical operator (NESO)~\cite{Hu:2021lnx,Tiwari:2024trl,She:2024rnx,She:2025qri}.
Related studies of spin dynamics under Bjorken and boost-invariant expansion can be found in Refs.~\cite{Wang:2021ngp,Drogosz:2024lkx,Singh:2026ytd}.
The stability and causality properties of second order spin hydrodynamics have also been investigated in Ref.~\cite{Daher:2024bah}.

Furthermore, a related development that has significantly influenced the modern understanding of relativistic hydrodynamics is the notion of the hydrodynamic attractor, brought to prominence by Heller and Spaliński in 2015 \cite{Heller:2015dha}. 
In this picture, the applicability of hydrodynamics is not tied solely to the convergence of the conventional gradient expansion or to the immediate establishment of local thermal equilibrium. 
Instead, it is associated with a dynamical loss of memory of the initial state, as transient non-hydrodynamic modes are damped and far-from-equilibrium solutions approach a common hydrodynamic trajectory.
Hydrodynamic attractors have been extensively studied in various expansion geometries, including Bjorken flow, Gubser flow, Hubble-type expansion, and nonconformal systems \cite{Heller:2015dha,Denicol:2016bjh,Heller:2021oxl,Heller:2016rtz,Chattopadhyay:2021ive,Jaiswal:2022mdk,Jaiswal:2022udf}, and in several microscopic and effective frameworks, such as kinetic theory, anisotropic hydrodynamics, and strongly coupled holographic systems \cite{Strickland:2017kux,Romatschke:2017vte,Denicol:2018pak,Du:2021fok,Romatschke:2017acs,Chen:2021wwh,Strickland:2018ayk,Almaalol:2020rnu,Denicol:2019lio}. 
The hydrodynamic attractor can also be viewed as a trajectory connecting the early time free streaming fixed point to the late time hydrodynamic fixed point, capturing the universal loss of initial condition memory \cite{Blaizot:2017ucy,Blaizot:2021cdv}.
This suggests that a phase space perspective, and in particular the analysis of fixed point structures, provides a useful complementary way to characterize hydrodynamization.
Other recent developments~\cite{Denicol:2017lxn,Behtash:2018moe,Behtash:2017wqg,Behtash:2019txb}, such as limiting attractors \cite{Boguslavski:2023jvg,Boguslavski2024LimitingAttractors}, adiabatic hydrodynamization \cite{Brewer:2019oha,Rajagopal:2024lou}, and phase space formulations \cite{Heller:2020anv,Spalinski:2025ngd,Du:2022bel}, have further connected attractor physics with fixed-point structures, pre-hydrodynamic slow modes, and dimensional reduction in phase space; see reviews \cite{Romatschke:2019,Soloviev:2021lhs,Jankowski:2023} and references therein.

Despite the extensive studies of hydrodynamic attractors in conventional relativistic hydrodynamics, their role in spin hydrodynamics remains much less explored. 
A recent work~\cite{Wang:2024afv} studied, for the first time, the attractor behavior of the decay rate of spin density in minimal causal spin hydrodynamics under Bjorken expansion. 
Their results suggest that, under certain conditions, spin density may behave as a conventional hydrodynamic variable at late times. 
In a closely related recent development, the analysis was extended to Gubser flow in Ref.~\cite{Li:2026vld}.
However, a systematic analysis of attractor behavior in $(0+1)$D Bjorken flow within relativistic spin hydrodynamics based on Zubarev's NESO formalism is still lacking. 
This motivates us to derive the Bjorken flow equations in this framework and to investigate the corresponding early and late time attractor behavior. 
Going beyond the minimal causal spin hydrodynamic model of \cite{Xie:2023gbo}, we formulate the attractor problem within a generalized second order spin hydrodynamic framework derived from Zubarev's NESO formalism based on~\cite{She:2024rnx,She:2025qri}.
This generalized NESO framework incorporates spin relaxation and spin vorticity coupling, together with dissipative, rotational stress, and boost heat vector sectors. The corresponding second-order terms in the relaxation constitutive equations include nonlocal memory corrections associated with two-point correlation functions and nonlinear response contributions associated with three-point correlation functions. This allows us to study how these nonlocal memory corrections and nonlinear response contributions influence the early and late time behavior.

Although spin is not a strict hydrodynamic variable, it can be regarded as a slow non-hydrodynamic mode that may be retained in an extended hydrodynamic description~\cite{Hongo:2021ona,Cao:2022aku,Becattini:2018duy,Becattini:2013fla,Wang:2024afv,Li:2026vld,Li:2020eon} 
Therefore, for finite spin relaxation, spin cannot support an independent hydrodynamic attractor in the strict late time limit. 
However, its early or intermediate time dynamics may still exhibit a transient attractor or a spin sector fixed point before spin relaxation becomes complete.

Another motivation arises from the poorly understood early time behavior of the spin density and its sensitivity to unknown initial conditions \cite{Hongo:2021ona,Becattini:2020ngo,Florkowski:2018fap,Wang:2024afv,Becattini:2022zvf}. 
Since spin polarization may retain information about the early stage evolution of the QGP, it is important to determine whether spin degrees of freedom lose memory of their initial conditions and approach universal non-equilibrium attractor behavior. 
Although the early time attractor in the minimal causal spin hydrodynamic framework was investigated numerically in Ref.~\cite{Wang:2024afv}, its fixed point structure and branch selection have not been systematically analyzed. 
In the present generalized second order framework, this kind of analysis allows us to identify the dynamically selected branches and the mechanisms controlling early time spin dynamics, assess the hydrodynamic relevance of spin degrees of freedom, and distinguish universal attractor behavior from model dependent features.

The paper is organized as follows.
Section~\ref{section2} provides a concise overview of relativistic spin hydrodynamics within the Zubarev's formalism for non-equilibrium statistical operators.
In Section~\ref{section3}, we derive the complete set of $(0+1)$D Bjorken-flow equations and the associated dimensionless attractor equation within the spin hydrodynamic formalism based on Zubarev's NESO.
Section~\ref{section4} analyzes the early and late time behavior of the attractors, and also presents numerical results.
Finally, Section~\ref{section5} summarizes and discusses our results.

\section{Review of Relativistic Spin Hydrodynamics}
\label{section2}

Relativistic spin hydrodynamics is a generalization of conventional relativistic hydrodynamics, aiming to describe the dynamics of relativistic fluids composed of particles with intrinsic spin angular momentum, such as quarks. Its core feature lies in the necessity to address the conservation and conversion of total angular momentum, comprising both orbital and spin contributions, in addition to the conservation laws of energy, momentum, and charge
\begin{align}
\partial_\mu N_a^{\mu} &= 0, \label{eq:consN}\\
\partial_\mu T^{\mu\nu} &= 0, \label{eq:consT}\\
\partial_\lambda J^{\lambda\mu\nu} &=0, \label{eq:consJ}
\end{align}
where the total angular momentum tensor $J^{\lambda\mu\nu}$ can be decomposed into an orbital part and a spin part
\begin{align}
J^{\lambda\mu\nu}
= x^\mu T^{\lambda\nu}-x^\nu T^{\lambda\mu}+S^{\lambda\mu\nu}.
\end{align}
Here, $T^{\mu\nu}$ is the energy-momentum tensor and $S^{\lambda\mu\nu}$ is the spin tensor.

Combining the conservation of total angular momentum with the conservation of energy and momentum, $\partial_\mu T^{\mu\nu}=0$, yields a crucial coupling equation between spin and orbital angular momentum
\begin{align}
\partial_\lambda S^{\lambda\mu\nu}+2T^{[\mu\nu]}=0.
\end{align}
This equation reveals a key physical insight: spin is strictly conserved only when the energy-momentum tensor $T^{\mu\nu}$ is symmetric. If $T^{\mu\nu}$ possesses an antisymmetric part, that antisymmetric part acts as a ``torque" that drives the conversion between spin and orbital angular momentum.

The role of pseudogauge transformations in spin hydrodynamics has also been discussed recently in Ref.~\cite{Montenegro:2026phf}. In the present work, we fix the pseudogauge freedom by adopting the totally antisymmetric spin tensor, which ensures thermodynamic consistency and yields a well-defined set of conserved currents. In the canonical‑like pseudogauge adopted in Refs.~\cite{She:2025qri,She:2024rnx}, we use the corresponding set of pseudogauge dependent conserved currents and decompose them into ideal and dissipative parts as
\begin{align}
T^{\mu\nu} &=
\epsilon\,u^\mu u^\nu - p\,\Delta^{\mu\nu}
+ h^\mu u^\nu + h^\nu u^\mu
+ \pi^{\mu\nu}
+ q^\mu u^\nu - q^\nu u^\mu
+ \phi^{\mu\nu}, \label{eq:Tdecomp}\\
N_a^\mu &= n_a u^\mu + j_a^\mu, \label{eq:Ndecomp}\\
S^{\lambda\mu\nu} &= u^\lambda S^{\mu\nu}+u^\mu S^{\nu\lambda}+u^\nu S^{\lambda\mu}+\varpi^{\lambda\mu\nu}, \label{eq:Sdecomp}
\end{align}
with $\Delta^{\mu\nu}=g^{\mu\nu}-u^\mu u^\nu$.
Here $u^\mu$ is the fluid four velocity, $\epsilon$ is the local energy density, and $p$ is the isotropic pressure. The vector $h^\mu$ denotes the energy-diffusion current, $\pi^{\mu\nu}$ the symmetric traceless shear stress tensor, $q^\mu$ the boost-heat vector associated with the antisymmetric sector of the energy-momentum tensor, and $\phi^{\mu\nu}$ the antisymmetric rotational stress tensor. In the charge sector, $n_a$ and $j_a^\mu$ denote the density and diffusion current of the conserved charge labelled by $a$. In the spin-current decomposition, $S^{\mu\nu}$ denotes the local spin density and $\varpi^{\lambda\mu\nu}$ the dissipative spin current. The latter is distinct from the spin chemical potential $\omega^{\mu\nu}$ introduced below, which is thermodynamically conjugate to $S^{\mu\nu}$.
We use
\begin{equation}
D\equiv u^\mu\partial_\mu,\qquad
\theta\equiv D_\mu u^\mu,\qquad
\nabla^\mu\equiv\Delta^{\mu\nu}\partial_\nu,
\end{equation}
where $D_\mu$ denotes the spacetime Levi--Civita covariant derivative. We use $\dot A\equiv DA$ for scalar quantities, while for dissipative vectors and tensors the dot denotes the corresponding projected comoving derivative.

The dissipative fluxes satisfy $u_\mu h^\mu=u_\mu j_a^\mu=u_\mu q^\mu=u_\mu\pi^{\mu\nu}=u_\mu\phi^{\mu\nu}=0$,
$\pi^\mu{}_\mu=0$, $\pi^{\mu\nu}=\pi^{\nu\mu}$, $\phi^{\mu\nu}=-\phi^{\nu\mu}$, and $\varpi^{\lambda\mu\nu}$ is totally antisymmetric and transverse in all indices.
The orthogonality condition in all indices is
\[
u_\lambda\varpi^{\lambda\mu\nu}
=u_\mu\varpi^{\lambda\mu\nu}
=u_\nu\varpi^{\lambda\mu\nu}=0.
\]
A single pair of angle brackets denotes projection orthogonal to the flow, e.g. $A^{\langle\mu\rangle}\equiv\Delta^\mu{}_{\nu}A^\nu$, while $A_{\langle\mu}B_{\nu\rangle}$ denotes the symmetric, transverse and traceless rank-two projection. Square brackets denote antisymmetrization.
The spin density satisfies the Frenkel condition $S^{\mu\nu}u_\mu=0$, and the spin chemical potential is taken to be transverse, $\omega_{\mu\nu}u^\mu=0$.
We adopt the power counting $\omega_{\mu\nu}\sim O(\partial)$.
The isotropic pressure appearing in the decomposition is not explicitly separated into equilibrium and bulk viscous channel. The bulk viscous pressure is defined as the deviation of the statistical average of the pressure operator from its equilibrium value
\begin{equation}
\Pi \equiv \langle \hat p \rangle - p(\epsilon,n_a,S_{\alpha\beta}) .
\end{equation}
In this expression, the hat denotes the corresponding microscopic operator and $\langle\cdots\rangle$ its expectation value in the nonequilibrium statistical ensemble, so that $\langle\hat p\rangle$ is the local nonequilibrium pressure.
The charge diffusion current corrected for energy flow is
\begin{equation}
\mathscr{J}_a^\mu
\equiv j_a^\mu-\frac{n_a}{h}h^\mu ,
\end{equation}
where $h=\epsilon+p$ is the enthalpy density, and $\mathscr{J}_a^\mu$ separates the charge diffusion contribution from the energy flow term $h^\mu$ in Eq.~\eqref{eq:Tdecomp}.

To account for spin degrees of freedom, we extend the standard local thermodynamics by treating $S^{\mu\nu}$ and $\omega_{\mu\nu}$ as a conjugate pair. The first law for local variables then takes the form
\begin{equation}
T\,\delta s
=\delta\epsilon-\sum_a \mu_a\,\delta n_a-\frac12\,\omega_{\alpha\beta}\,\delta S^{\alpha\beta}. \label{eq:firstlaw}
\end{equation}
Defining the local pressure from the equilibrium thermodynamic potential yields the generalized Euler relation
\begin{align}
\epsilon+p
=Ts+\sum_a \mu_a n_a+\frac12\,\omega_{\alpha\beta}S^{\alpha\beta}= h, \label{eq:enthalpy}
\end{align}
and the corresponding Gibbs–Duhem relation
\begin{align}
\delta p
=s\,\delta T+\sum_a n_a\,\delta\mu_a+\frac12\,S^{\alpha\beta}\,\delta\omega_{\alpha\beta}. \label{eq:gibbsduhem}
\end{align}
For the constitutive relations below, we introduce the inverse temperature $\beta\equiv 1/T$ and the dimensionless charge chemical potentials $\alpha_a\equiv\beta\mu_a$. The dimensionful variables $\mu_a$ are retained in the local thermodynamic relations, whereas $\alpha_a$ enter the thermodynamic forces in the constitutive relations. Similarly, we use the dimensionful spin chemical potential $\omega^{\mu\nu}$ and its dimensionless counterpart $\Omega^{\mu\nu}\equiv\beta\omega^{\mu\nu}$, with the former entering the spin thermodynamic relations and the latter the NESO thermodynamic forces.

Following Refs.~\cite{She:2025qri,She:2024rnx}, the thermodynamic forces and second order structures used below, such as
$M_\mu$, $\xi_{\mu\nu}$, $\Gamma$, $\delta_a$,
$\mathcal X$, $\mathcal Y_a$, $\mathcal Z^{\alpha\beta}$,
$\mathcal H_\mu$, $\mathcal Q_\mu$, $\mathcal K_{\alpha\beta}$,
$\mathcal R_{\mu\nu}$, and $\varXi^{\lambda\mu\nu}$,
are defined as in those references.
The relaxation equations for the dissipative currents in second‑order spin hydrodynamics, obtained via intricate tensor calculations, feature on the left‑hand side the dissipative current and its relaxation‑time term, while the right‑hand side comprises the first‑order Navier–Stokes terms and second‑order gradient corrections. These second‑order contributions stem from nonlocal effects represented by two‑point correlation functions, from two‑point functions of the generalized thermodynamic forces, and from nonlinear corrections due to three‑point correlation functions.
The multiple subscripts of the transport coefficients label the corresponding coupled channels rather than Lorentz tensor indices. Only the combinations relevant to the Bjorken and spin probe reduction will be used explicitly in Sec.~\ref{section3}.

The Israel--Stewart-type relaxation equations, together with the second order constitutive relation for $\varpi^{\lambda\mu\nu}$, read
\begin{align}
\pi_{\mu\nu}+\tau_\pi\Delta_{\mu\nu\rho\sigma}D\pi^{\rho\sigma}
=&\ 2\eta\sigma_{\mu\nu}
+\widetilde{\eta}_{\pi}\theta\pi_{\mu\nu}
+2\widetilde{\eta}\theta\Gamma\sigma_{\mu\nu}
+2\widetilde{\eta}Th^{-1}\sum_{a}n_{a}\dot{u}_{\langle\mu}\nabla_{\nu\rangle}\alpha_{a}
\notag\\
&+2\eta_{\pi p\pi}\theta\sigma_{\mu\nu}
+\sum_{ab}\eta_{\pi\mathscr{J}_{a}\mathscr{J}_{b}}
\nabla_{\langle\mu}\alpha_{a}\nabla_{\nu\rangle}\alpha_{b}
+2\sum_{a}\eta_{\pi\mathscr{J}_{a}q}
\nabla_{\langle\mu}\alpha_{a}M_{\nu\rangle}
\notag\\
&+\eta_{\pi qq}M_{\langle\mu}M_{\nu\rangle}
+\eta_{\pi\pi\pi}\sigma_{\alpha\langle\mu}\sigma_{\nu\rangle}^{\,\,\,\,\alpha}
+2\eta_{\pi\pi\phi}\sigma_{\alpha\langle\mu}\xi_{\nu\rangle}^{\,\,\,\,\alpha}
+\eta_{\pi\phi\phi}\xi_{\alpha\langle\mu}\xi_{\nu\rangle}^{\,\,\,\,\alpha},
\end{align}
\begin{align}
\Pi+\tau_{\Pi}D\Pi
=&\ -\zeta\theta
+\widetilde{\zeta}_{\Pi}\theta\Pi
+\left[
\frac{1}{2}\frac{\partial^{2}p}{\partial\epsilon^{2}}\zeta_{\epsilon p}^{2}
+\frac{1}{2}\sum_{ab}\frac{\partial^{2}p}{\partial n_{a}\partial n_{b}}
\zeta_{n_{a}p}\zeta_{n_{b}p}
+\sum_{a}\frac{\partial^{2}p}{\partial\epsilon\partial n_{a}}
\zeta_{\epsilon p}\zeta_{n_{a}p}
\right]\theta^{2} \nonumber\\
&+\zeta_{S\phi}\mathcal{K}_{\alpha\beta}\xi^{\alpha\beta}
\notag -\left[
\Gamma\widetilde{\zeta}
+\widetilde{\Gamma}\widetilde{\zeta}_{p\epsilon}
+\sum_{a}\widetilde{\zeta}_{pn_{a}}\widetilde{\delta}_{a}
\right]\theta^{2}
-\sum_{a}\overline{\zeta}_{a}\dot{u}_{\rho}\nabla^{\rho}\alpha_{a}
\notag\\
&+\sum_{i}\zeta_{p\mathfrak{D}_{i}}
\left[
\left(\partial_{\epsilon n}^{i}\beta\right)\mathcal{X}
+\sum_{a}\left(\partial_{\epsilon n}^{i}\alpha_{a}\right)\mathcal{Y}_{a}
+\left(\partial_{\epsilon n}^{i}\Omega_{\alpha\beta}\right)\mathcal{Z}^{\alpha\beta}
\right]
\notag\\
&+\zeta_{ppp}\theta^{2}
+\sum_{ab}\zeta_{p\mathscr{J}_{a}\mathscr{J}_{b}}
\nabla_{\alpha}\alpha_{a}\nabla^{\alpha}\alpha_{b}
+2\sum_{a}\zeta_{p\mathscr{J}_{a}q}\nabla^{\sigma}\alpha_{a}M_{\sigma}
+\zeta_{pqq}M^{\sigma}M_{\sigma}\nonumber\\
&
+\zeta_{p\pi\pi}\sigma^{\rho\sigma}\sigma_{\rho\sigma}
+\zeta_{p\phi\phi}\xi^{\rho\sigma}\xi_{\rho\sigma},
\end{align}
\begin{align}
\mathscr{J}_{c\mu}+\sum_{b}\tau_{\mathscr{J}}^{cb}\Delta_{\mu\beta}D\mathscr{J}_{b}^{\beta}
=&\ \sum_{b}\chi_{cb}\nabla_{\mu}\alpha_{b}
+\chi_{\mathscr{J}_{c}q}M_{\mu}
+\widetilde{\chi}_{\mathscr{J}_{c}h}h^{-2}
\left(\Gamma h+\sum_{b}\delta_{b}n_{b}\right)
\theta\sum_{a}n_{a}\nabla_{\mu}\alpha_{a}
\notag\\
&+\overline{\chi}^{c}\theta M_{\mu}
+\sum_{b}\tau_{\mathscr{J}}^{cb}\chi_{\mathscr{J}_{b}q}
\Delta_{\mu\beta}DM^{\beta}
+\sum_{b}\widetilde{\chi}_{\mathscr{J}}^{cb}\theta\mathscr{J}_{b\mu}
-\sum_{b}\widetilde{\chi}_{\mathscr{J}}^{cb}
\chi_{\mathscr{J}_{b}q}\theta M_{\mu}
\notag\\
&+\widetilde{\chi}_{\mathscr{J}_{c}q}\theta\Gamma M_{\mu}
+\widetilde{\chi}_{\mathscr{J}_{c}q}\Delta_{\mu\beta}DM^{\beta}
-\beta\theta\dot{u}_{\mu}\sum_{a}\delta_{a}\widetilde{\chi}_{ac}
\notag\\
&-\beta\widetilde{\chi}_{\mathscr{J}_{c}h}
\left(2\sigma_{\mu\nu}\dot{u}^{\nu}+y\theta\dot{u}_{\mu}\right)
-2\widetilde{\chi}_{\mathscr{J}_{c}q}\xi_{\mu\nu}\dot{u}^{\nu}
+\beta\chi_{\mathscr{J}_{c}h}\mathcal{H}_{\mu}
+\chi_{\mathscr{J}_{c}q}\mathcal{Q}_{\mu}
\notag\\
&+2\sum_{a}\chi_{\mathscr{J}_{c}p\mathscr{J}_{a}}\theta\nabla_{\mu}\alpha_{a}
+2\chi_{\mathscr{J}_{c}pq}\theta M_{\mu}
+2\sum_{a}\chi_{\mathscr{J}_{c}\mathscr{J}_{a}\pi}
\nabla^{\nu}\alpha_{a}\sigma_{\mu\nu}
\notag\\
&+2\sum_{a}\chi_{\mathscr{J}_{c}\mathscr{J}_{a}\phi}
\nabla^{\nu}\alpha_{a}\xi_{\mu\nu}
+2\chi_{\mathscr{J}_{c}q\pi}M^{\nu}\sigma_{\mu\nu}
+2\chi_{\mathscr{J}_{c}q\phi}M^{\nu}\xi_{\mu\nu},
\end{align}
\begin{align}
\phi_{\mu\nu}+\tau_{\phi}\mydelta_{\mu\nu\rho\sigma}D\phi^{\rho\sigma}
=&\ 2\gamma\xi_{\mu\nu}
+\widetilde{\gamma}_{\phi}\theta\phi_{\mu\nu}
+2\widetilde{\gamma}\theta\Gamma\xi_{\mu\nu}
+2\widetilde{\gamma}Th^{-1}\sum_{a}n_{a}
\dot{u}_{[\mu}\nabla_{\nu]}\alpha_{a}
+\gamma_{\phi S}\mathcal{R}_{\langle\mu\rangle\langle\nu\rangle}
\notag\\
&+2\gamma_{\phi p\phi}\theta\xi_{\mu\nu}
+\sum_{ab}\gamma_{\phi\mathscr{J}_{a}\mathscr{J}_{b}}
\nabla_{[\mu}\alpha_{a}\nabla_{\nu]}\alpha_{b}
+2\sum_{a}\gamma_{\phi\mathscr{J}_{a}q}
\nabla_{[\mu}\alpha_{a}M_{\nu]}
\notag\\
&+\gamma_{\phi\pi\pi}\sigma_{\alpha[\mu}\sigma_{\nu]}^{\,\,\,\,\alpha}
+2\gamma_{\phi\pi\phi}\sigma_{\alpha[\mu}\xi_{\nu]}^{\,\,\,\,\alpha}
+\gamma_{\phi\phi\phi}\xi_{\alpha[\mu}\xi_{\nu]}^{\,\,\,\,\alpha},
\label{relaxation_phi}
\end{align}
\begin{align}
q_{\mu}+\tau_{q}\Delta_{\mu\nu}Dq^{\nu}
=&\ \sum_{a}\lambda_{q\mathscr{J}_{a}}\nabla_{\mu}\alpha_{a}
-\lambda M_{\mu}
+\widetilde{\lambda}_{qh}h^{-2}
\left(\Gamma h+\sum_{c}\delta_{c}n_{c}\right)
\theta\sum_{a}n_{a}\nabla_{\mu}\alpha_{a}\nonumber\\
&+\sum_{a}\widetilde{\lambda}_{q\mathscr{J}_{a}}
\Delta_{\mu\gamma}D\left(\nabla^{\gamma}\alpha_{a}\right)
-\widetilde{\lambda}\theta\Gamma M_{\mu}
+\tau_{q}\Delta_{\mu\gamma}\sum_{a}\lambda_{q\mathscr{J}_{a}}
D\left(\nabla^{\gamma}\alpha_{a}\right)
\notag\\
&+\sum_{a}\overline{\lambda}^{a}\theta\nabla_{\mu}\alpha_{a}
+\widetilde{\lambda}_{q}\theta q_{\mu}
-\sum_{a}\widetilde{\lambda}_{q}\lambda_{q\mathscr{J}_{a}}
\theta\nabla_{\mu}\alpha_{a}
-2\widetilde{\lambda}_{qh}\beta\sigma_{\mu\nu}\dot{u}^{\nu}
-\widetilde{\lambda}_{qh}y\beta\theta\dot{u}_{\mu}
\notag\\
&+2\widetilde{\lambda}\xi_{\mu\nu}\dot{u}^{\nu}
-\sum_{a}\widetilde{\lambda}_{q\mathscr{J}_{a}}
\beta\theta\dot{u}_{\mu}\delta_{a}
+\lambda_{qh}\mathcal{H}_{\mu}
-\lambda\mathcal{Q}_{\mu}
+2\sum_{a}\lambda_{qp\mathscr{J}_{a}}\theta\nabla_{\mu}\alpha_{a}
\notag\\
&+2\lambda_{qpq}\theta M_{\mu}
+2\sum_{a}\lambda_{q\mathscr{J}_{a}\pi}
\nabla^{\nu}\alpha_{a}\sigma_{\mu\nu}
+2\sum_{a}\lambda_{q\mathscr{J}_{a}\phi}
\nabla^{\nu}\alpha_{a}\xi_{\mu\nu}
+2\lambda_{qq\pi}M^{\nu}\sigma_{\mu\nu}
\notag\\
&+2\lambda_{qq\phi}M^{\nu}\xi_{\mu\nu},
\end{align}
\begin{align}
\varpi^{\lambda\mu\nu}=\ \varphi\varXi^{\lambda\mu\nu}+2\sum_{a}\varphi_{\varpi\mathscr{J}_{a}\phi}\myDelta^{\lambda\mu\nu\rho\sigma\delta}\xi_{\sigma\delta}\nabla_{\rho}\alpha_{a}+2\varphi_{\varpi q\phi}\myDelta^{\lambda\mu\nu\rho\sigma\delta}M_{\rho}\xi_{\sigma\delta},
\end{align}
where we also define projectors orthogonal to $u^\mu$ through
\begin{align}
\Delta_{\rho\sigma}^{\mu\nu}	=&\frac{1}{2}\left(\Delta_{\rho}^{\mu}\Delta_{\sigma}^{\nu}+\Delta_{\sigma}^{\mu}\Delta_{\rho}^{\nu}\right)-\frac{1}{3}\Delta^{\mu\nu}\Delta_{\rho\sigma},\\
\mydelta_{\rho\sigma}^{\mu\nu}	=&\frac{1}{2}\left(\Delta_{\rho}^{\mu}\Delta_{\sigma}^{\nu}-\Delta_{\sigma}^{\mu}\Delta_{\rho}^{\nu}\right),\\
\myDelta_{\alpha\beta\gamma}^{\lambda\mu\nu}	=&\frac{1}{6}\left(\Delta_{\alpha}^{\lambda}\Delta_{\beta}^{\mu}\Delta_{\gamma}^{\nu}+\Delta_{\alpha}^{\mu}\Delta_{\beta}^{\nu}\Delta_{\gamma}^{\lambda}+\Delta_{\alpha}^{\nu}\Delta_{\beta}^{\lambda}\Delta_{\gamma}^{\mu}-\Delta_{\alpha}^{\mu}\Delta_{\beta}^{\lambda}\Delta_{\gamma}^{\nu}-\Delta_{\alpha}^{\nu}\Delta_{\beta}^{\mu}\Delta_{\gamma}^{\lambda}-\Delta_{\alpha}^{\lambda}\Delta_{\beta}^{\nu}\Delta_{\gamma}^{\mu}\right).
\end{align}
In this formulation, the dissipative fluid-dynamical currents appear as independent dynamical variables, which relax to the values of Navier-Stokes theory on characteristic time scales given by the relaxation times $\tau_\pi,\tau_\Pi,\tau^{cb}_\mathscr{J},\tau_\phi$, and $\tau_q$. This type of formalism was referred to as transient theory of fluid dynamics, since it describes this relaxation (or transient) process of each dissipative current toward its respective (asymptotic) Navier-Stokes value.

Setting all relaxation times and second order coefficients to zero gives the first-order linear relations
\begin{align}
\pi_{\mu\nu} &= 2\eta\sigma_{\mu\nu}, \\
\Pi &= -\zeta\theta, \\
\mathscr{J}_{c\mu} &= \sum_b\chi_{cb}\nabla_\mu\alpha_b
+\chi_{\mathscr{J}_cq}M_\mu, \\
\phi_{\mu\nu} &= 2\gamma\xi_{\mu\nu}, \\
q_\mu &= \sum_a\lambda_{q\mathscr{J}_a}\nabla_\mu\alpha_a-\lambda M_\mu, \\
\varpi^{\lambda\mu\nu} &= \varphi\varXi^{\lambda\mu\nu}.
\end{align}
When the spin force and spin-sector dissipative variables are switched off, the conventional charge, shear, bulk, and heat-flow sectors are recovered. In Sec.~\ref{section3}, after imposing charge neutrality and Bjorken symmetry, the vector charge structures and the dissipative spin flux drop out of the $xy$ spin channel, leaving the rotational stress relaxation equation used in the attractor analysis.

\section{Bjorken Flow in Spin Hydrodynamics Based on Zubarev's NESO Formalism}
\label{section3}

\subsection{Bjorken Flow}
In this section we specialize the general hydrodynamic framework developed above to the Bjorken flow setup. Before imposing the spacetime symmetries, we restrict ourselves to the sector of vanishing conserved charges densities and chemical potentials
\begin{equation}
n_a=\mu_a=\alpha_a=j_a^\mu=\mathscr{J}_a^\mu=0 .
\end{equation}
In this sector the conserved charge dynamics decouples from the energy momentum dynamics. (Equivalently, one may first impose Bjorken symmetry on the full hydrodynamic equations and then take the limits, which gives the same reduced system.)

We specialize to $(0+1)$D Bjorken flow in Milne coordinates
$x^{\mu}=(\tau\equiv\sqrt{t^{2}-z^{2}},x,y,\eta_{s}\equiv\frac{1}{2}\ln\frac{t+z}{t-z})$. The line element and metric are
\[
ds^2=d\tau^2-dx^2-dy^2-\tau^2d\eta_s^2,
\qquad
g_{\mu\nu}=\mathrm{diag}(1,-1,-1,-\tau^2).
\]
After changing from Cartesian to Milne coordinates, spacetime divergences are understood as Levi--Civita covariant divergences. In particular, for Bjorken flow
\begin{equation}
u^{\mu}=(1,0,0,0),\qquad D=\partial_{\tau},\qquad\theta=D_{\mu}u^{\mu}=\frac{1}{\sqrt{-g}}\partial_{\mu}\!\left(\sqrt{-g}\,u^{\mu}\right)=\frac{1}{\tau},
\end{equation}
where $D_{\mu}$ denotes the spacetime covariant derivative, and $g\equiv \det(g_{\mu\nu})$ is the determinant of the metric tensor. For $(0+1)$D Bjorken flow, all hydrodynamic fields depend only on the proper time $\tau$. 
The Bjorken symmetries forbid vector dissipative currents, and in the Landau frame the energy diffusion current vanishes.
Thus,
\begin{equation}
h^\mu=0,\qquad q^\mu=0 .
\end{equation}

Inserting Eq.~\eqref{eq:Sdecomp} to Eq.~\eqref{eq:consJ}, we derive the spin conservation equation as
\begin{align}
&D S^{\mu\nu}
+\theta S^{\mu\nu}
+u^{\mu}\nabla_{\lambda}S^{\nu\lambda}
+S^{\nu\lambda}\nabla_{\lambda}u^{\mu}
+u^{\nu}\nabla_{\lambda}S^{\lambda\mu}
+S^{\lambda\mu}\nabla_{\lambda}u^{\nu}
+\nabla_{\lambda}\varpi^{\lambda\mu\nu}
\nonumber\\
&\hspace{4cm}
+2q^{\mu}u^{\nu}
-2q^{\nu}u^{\mu}
+2\phi^{\mu\nu}
=0 .
\end{align}
The Frenkel condition,
\begin{equation}
S^{\mu\nu}u_\mu=0 ,
\end{equation}
constrains the spin density tensor to be purely spatial in the local rest frame of the fluid. The independent components are therefore
\begin{equation}
S^{xy},\qquad S^{x\eta},\qquad S^{y\eta}.
\end{equation}
We focus on the longitudinal spin polarization sector. 
Under transverse rotations, ($S^{x\eta},S^{y\eta}$) transforms as a transverse vector and is forbidden by transverse rotational invariance
\begin{equation}
S^{x\eta}=S^{y\eta}=0 ,
\end{equation}
so that the only retained spin density component is
\begin{equation}
S\equiv S^{xy}.
\end{equation}
The same conclusion that \(S^{xy}\) is the only retained spin density component in a Bjorken symmetric setup can also be found in Ref.~\cite{Wang:2024afv}.

Taking the $xy$ component of the spin conservation equation, using Bjorken symmetry and the transversality of the dissipative spin current,
\begin{equation}
u_\lambda\varpi^{\lambda\mu\nu}=0 ,
\end{equation}
one finds
\begin{equation}
\nabla_\lambda\varpi^{\lambda xy}=0 .
\end{equation}
This means that, under the symmetry reduction, the dissipative spin flux does not contribute to the reduced $xy$ spin equation. 

However, the spin density can still evolve locally through spin orbital exchange mediated by the antisymmetric rotational tensor.
The spin conservation law then reduces to
\begin{equation}
\partial_\tau S+\frac{S}{\tau}+2\phi=0 ,
\label{eq:spin_bjorken}
\end{equation}
where
\begin{equation}
\phi\equiv\phi^{xy}.
\end{equation}
The same conclusion as Eq.(\ref{eq:spin_bjorken}) can also be found in Appendix A of Ref.~\cite{Wang:2024afv}.

We now reduce the second order constitutive relation for the rotational stress tensor,
Eq.~\eqref{relaxation_phi}. 
In the present setup, all charge diffusion structures are absent, the following related terms vanished
\begin{equation}
2\widetilde{\gamma}Th^{-1}\sum_a n_a
\dot{u}_{[\mu}\nabla_{\nu]}\alpha_a,\qquad
\sum_{ab}\gamma_{\phi\mathscr{J}_a\mathscr{J}_b}
\nabla_{[\mu}\alpha_a\nabla_{\nu]}\alpha_b,\qquad
2\sum_a\gamma_{\phi\mathscr{J}_a q}
\nabla_{[\mu}\alpha_a M_{\nu]}.
\end{equation}
The shear related second order structures and nonlinear terms involving the spin force $\xi_{\mu\nu}$ are also not retained in the probe limit attractor analysis.
In addition, in the \(xy\) channel, the Bjorken shear tensor is diagonal, giving
\begin{equation}
\sigma_{\alpha[x}\sigma_{y]}^{\,\,\,\,\alpha}=0 ,
\end{equation}
and the single channel spin force structure satisfies
\begin{equation}
\xi_{\alpha[x}\xi_{y]}^{\,\,\,\,\alpha}=0 .
\end{equation}
Thus, the terms
\begin{equation}
\gamma_{\phi\pi\pi}\sigma_{\alpha[\mu}\sigma_{\nu]}^{\,\,\,\,\alpha},
\qquad
2\gamma_{\phi\pi\phi}\sigma_{\alpha[\mu}\xi_{\nu]}^{\,\,\,\,\alpha},
\qquad
\gamma_{\phi\phi\phi}\xi_{\alpha[\mu}\xi_{\nu]}^{\,\,\,\,\alpha}    
\end{equation}
are not kept in the reduced equation.

The projected generalized thermodynamic force is defined as in Refs.~\cite{She:2025qri,She:2024rnx},
\begin{equation}
\mathcal{R}_{\langle\mu\rangle\langle\nu\rangle}
\equiv
\Delta_{\mu}{}^{\alpha}\Delta_{\nu}{}^{\beta}
\mathcal{R}_{\alpha\beta},
\qquad
\mathcal{R}_{\mu\nu}
=
\theta\mathcal{K}_{\mu\nu}
+\mathcal{W}_{\mu\nu},
\end{equation}
where
\begin{equation}
\mathcal{W}_{\mu\nu}
\equiv
-\frac{1}{2\beta}
\left(
\Delta_{\mu}{}^{\rho}\Delta_{\nu}{}^{\sigma}
-\Delta_{\nu}{}^{\rho}\Delta_{\mu}{}^{\sigma}
\right)
u^{\alpha}\partial_{\sigma}\Omega_{\alpha\rho}.
\end{equation} The spin dependence of the equation of state is neglected in the probe limit, so that
\begin{equation}
\mathcal{K}_{\mu\nu}
=
\frac{\partial p}{\partial S^{\mu\nu}}
\Big|_{\epsilon,n_a}
=0 .
\end{equation}
For the $(0+1)$D Bjorken flow, the hydrodynamic variables depend only on $\tau$, and therefore the projected spatial gradients vanish. Hence
\begin{equation}
\mathcal{W}_{\mu\nu}=0,
\qquad
\mathcal{R}_{\langle\mu\rangle\langle\nu\rangle}=0 .
\end{equation}
Therefore, the second order generalized thermodynamic force term
\begin{equation}
\gamma_{\phi S}\mathcal{R}_{\langle\mu\rangle\langle\nu\rangle}
\end{equation}
also vanishes.
Finally, the second order constitutive relation for the rotational stress tensor reduces to
\begin{align}
\phi_{\mu\nu}+\tau_{\phi}\dot{\phi}_{\mu\nu}
= 2\gamma\xi_{\mu\nu}
+\widetilde{\gamma}_{\phi}\theta\phi_{\mu\nu}
+2\widetilde{\gamma}\theta\Gamma\xi_{\mu\nu} 
+2\gamma_{\phi p\phi}\theta\xi_{\mu\nu}.
\label{eq:constitutive_relation_ori}
\end{align}

The two source terms have the same tensor structure.
For the attractor analysis, it is therefore convenient to combine them into a single effective source correction,
\begin{equation}
2\widetilde{\gamma}\theta\Gamma\xi_{\mu\nu}
+
2\gamma_{\phi p\phi}\theta\xi_{\mu\nu}
=
2\widetilde{\gamma}_p\Gamma\theta\xi_{\mu\nu}.
\end{equation}
This defines
\begin{equation}
\widetilde{\gamma}_p
\equiv
\widetilde{\gamma}
+
\frac{\gamma_{\phi p\phi}}{\Gamma}.
\end{equation}
Here \(\Gamma\) denotes the thermodynamic derivative, we have in the conformal limit, 
\begin{equation}
\Gamma
\equiv \left. \frac{\partial p}{\partial \epsilon}\right|_{n_a,S^{\mu\nu}} =\frac{1}{3}.
\end{equation}
The constitutive relation used in the following analysis is
\begin{equation}
\phi_{\mu\nu}+\tau_\phi\dot{\phi}_{\mu\nu}
=
2\gamma\xi_{\mu\nu}
+\widetilde{\gamma}_{\phi}\theta\phi_{\mu\nu}
+2\widetilde{\gamma}_p\Gamma\theta\xi_{\mu\nu}.
\label{eq:constitutive_reduced}
\end{equation}
The three coefficients in~Eq.(\ref{eq:constitutive_relation_ori}) that survive in the reduced rotational sector have distinct dynamical roles. The coefficient $\gamma$ multiplies the leading spin force $\xi_{\mu\nu}$ and controls the ordinary dissipative source response. The coefficient $\widetilde{\gamma}_\phi$ multiplies $\theta\phi_{\mu\nu}$; because this correction is proportional to the rotational stress itself, it changes the evolution through direct feedback and will be referred to as the ``self feedback'' coupling. By contrast, the terms $2\widetilde{\gamma}\theta\Gamma\xi_{\mu\nu}$ and $2\gamma_{\phi p\phi}\theta\xi_{\mu\nu}$ are proportional to the thermodynamic spin force rather than to $\phi_{\mu\nu}$ and therefore act as expansion induced source corrections. Microscopically, the self feedback correction originates from nonlocal memory effects, while the source corrections contain both nonlocal memory and nonlinear response contributions~\cite{She:2025qri,She:2024rnx}.

The antisymmetric thermodynamic force conjugate to the rotational stress tensor is defined in Refs.~\cite{She:2025qri,She:2024rnx} as
\begin{align}
    \xi_{\mu\nu}
    =
    \Delta^{\lambda\delta}_{\mu\nu}
    \left(
    \partial_{\lambda}u_{\delta}
    +
    \beta^{-1}\Omega_{\lambda\delta}
    \right),
\end{align}
where \(\Omega_{\mu\nu}=\beta\omega_{\mu\nu}\). 
In the \((0+1)\) dimensional Bjorken flow, the velocity gradient contribution to the \(xy\) component vanishes. Therefore,
\begin{align}
    \xi^{xy}
    =
    \beta^{-1}\Omega^{xy}
    =
    \omega^{xy}.
\end{align} 
Following Refs.~\cite{Hattori:2019lfp,Daher:2024ixz} and its subsequent formulation within the NESO framework~\cite{She:2025qri,She:2024rnx}, the system can be closed by the linear spin equation of state with respect to the dimensionful spin chemical potential,
\begin{align}
S_{\mu\nu}
=
\chi_{\omega} \omega_{\mu\nu}
+
2\chi' u_{\lambda}\omega^{\lambda[\mu}u^{\nu]} .
\end{align}
This is the general linear decomposition of the antisymmetric spin density in an isotropic medium when the fluid velocity $u^\mu$ is available as an additional vector. The first term gives the transverse spin susceptibility $\chi_\omega$, whereas the second term, with susceptibility $\chi'$, parametrizes the part involving components of the spin chemical potential along the fluid velocity. Under the Frenkel condition used here, this second structure does not contribute to the purely spatial $xy$ channel.
The susceptibility \(\chi_{\omega}\) is related to the susceptibility defined with respect to the dimensionless spin potential \(\Omega_{\mu\nu}\) by
\begin{align}
    \chi_{\omega}=\beta\chi_{\Omega}.
\end{align}
In local equilibrium with the Frenkel condition, the second term does not contribute to the longitudinal \(xy\) channel. Hence,
\begin{align}
S^{xy}
=
\chi_{\omega} \omega^{xy},
\end{align}
and one obtains
\begin{align}
    \xi^{xy}
    =
    \frac{S^{xy}}{\chi_{\omega}} = \frac{S}{\chi_{\omega}}.
\end{align}

Eq.\eqref{eq:spin_bjorken} and Eq.\eqref{eq:constitutive_reduced} then give
\begin{align}
&\partial_\tau S+\frac{S}{\tau}+2\phi=0,
\label{eq:spin_reduced_1}
\\
&\phi+\tau_\phi\partial_\tau\phi
=
2\gamma\frac{S}{\chi_{\omega}}
+\frac{2\Gamma\widetilde{\gamma}_p}{\tau}\frac{S}{\chi_{\omega}}
+\frac{\widetilde{\gamma}_{\phi}}{\tau}\phi .
\label{eq:spin_reduced_2}
\end{align}
Our derivation above is performed within the power counting scheme $\omega_{\mu\nu}\sim O(\partial^1)$.
Ref.~\cite{She:2025qri} gives the same result, Eq.~\eqref{eq:spin_reduced_2}, as Ref.~\cite{She:2024rnx,Tiwari:2024trl} with both the corresponding counting scheme and the power counting scheme $\omega_{\mu\nu}\sim O(\partial^0)$.

Finally, eliminating $\phi$ from Eqs.~\eqref{eq:spin_reduced_1} and \eqref{eq:spin_reduced_2} gives the second order equation
\begin{align}
\tau_\phi^2\frac{d^2S}{d\tau^2}
&+
\left(
1+\frac{\tau_\phi}{\tau}
-\frac{\widetilde{\gamma}_{\phi}}{\tau}
\right)
\tau_\phi\frac{dS}{d\tau}+
\left[
\frac{\tau_\phi}{\tau}
-\left(\frac{\tau_\phi}{\tau}\right)^2
+4\frac{\tau_\phi\gamma}{\chi_{\omega}}
+4\Gamma\frac{\tau_{\phi}\widetilde{\gamma}_p}{\tau \chi_{\omega}}
-\frac{\widetilde{\gamma}_{\phi}\tau_\phi}{\tau^2}
\right]S
=0 .
\label{eq:master_S_equation}
\end{align}

\subsection{Nondimensionalization}

The Knudsen number $K_n$ may be estimated as the ratio between the microscopic relaxation scale $\lambda_{\mathrm{mfp}}$ and the macroscopic expansion scale $L$,
\begin{align}
    K_n \equiv \frac{\lambda_{\mathrm{mfp}}}{L}
    \sim \frac{\tau_\phi}{\tau}
    \equiv w^{-1},
\end{align}
which motivates the introduction of the dimensionless variable
\begin{align}
    w \equiv \frac{\tau}{\tau_\phi}.
\end{align}
The hydrodynamic regime corresponds to $K_n\ll 1$, or equivalently $w\gg 1$ ~\cite{Heller:2015dha,Soloviev:2021lhs}.

In what follows, we work in the conformal limit, where the relaxation time is assumed to scale with the inverse temperature,
\begin{align}
    \tau_\phi(\tau)=c_\phi\,T^{-1}(\tau),
\end{align}
with $c_\phi$ a dimensionless constant. 
The $\tau$ and $\tau_\phi$ denote the dimensionless quantities obtained by measuring both times in units of a fixed inverse-temperature scale $T_0^{-1}$, i.e., $\tau\to T_0\tau$ and $\tau_\phi\to T_0\tau_\phi$. We use the same symbols for the rescaled quantities.
\begin{align}
    \tau_\phi=c_\phi\,\tau^{1/3},
    \qquad
    w=\frac{\tau}{\tau_\phi}.
\end{align}
Therefore,
\begin{align}
    \tau=(c_\phi w)^{3/2},
    \qquad
    \tau_\phi=c_\phi^{3/2}w^{1/2}.
\end{align}
Since the attractor analysis depends only on the scaling with the conformal clock variable $w$, the constant $c_\phi$ fixes the overall normalization of the dimensionless units.

Following Ref.~\cite{Wang:2024afv}, we introduce the relative decay rate of the spin density as
\begin{equation}
f(w) \equiv \frac{2w}{3S}\frac{dS}{dw}.
\label{eq:f_definition}
\end{equation}
Here the factor \(2/3\) follows from the conformal scaling
\(w\propto \tau^{2/3}\).
In terms of $f(w)$, the master equation Eq.~\eqref{eq:master_S_equation} rescales to
\begin{align}
\frac{2}{3}\frac{f'}{w}
+\frac{f^2}{w^2}
+\frac{f}{w}
-\frac{\kappa}{w^{2}} f
+
\left[
w^{-1}
-w^{-2}
+4\lambda 
+\frac{4\lambda \lambda_p}{3} w^{-1}  
-\frac{\kappa}{w^{2}}
\right]
=0.
\label{eq:master_attractor}
\end{align}
We parameterize the relevant transport coefficients through the following dimensionless combinations:
\begin{equation}
\lambda\equiv \frac{\tau_\phi\gamma}{\chi_{\omega}},
\qquad
\kappa\equiv \frac{\widetilde{\gamma}_\phi}{\tau_\phi},
\qquad
\lambda_p \equiv \frac{\widetilde{\gamma}_p}{\tau_\phi\gamma}.
\end{equation}
Using $w=\tau/\tau_\phi$, the corresponding source term can be written as
\begin{align}
4\Gamma\frac{\tau_\phi\widetilde{\gamma}_p}{\tau\chi_\omega}
&=
4\Gamma
\left(\frac{\tau_\phi\gamma}{\chi_\omega}\right)
\left(\frac{\widetilde{\gamma}_p}{\tau_\phi\gamma}\right)
\left(\frac{\tau_\phi}{\tau}\right)
=
\frac{4\lambda\lambda_p}{3}w^{-1},
\end{align}
where the factor $1/3$ originates explicitly from the conformal value $\Gamma=1/3$.

In the conformal limit, the relevant scalings are
\begin{align}
    \gamma\sim T^3,
    \qquad
    \chi_{\omega}\sim T^2,
    \qquad
    \widetilde{\gamma}_p\sim T^2,
    \qquad
    \widetilde{\gamma}_{\phi}\sim T^{-1}.
\end{align}
Therefore, \(\lambda\), \(\kappa\), and \(\lambda_p\) are dimensionless constants. This parametrization absorbs the microscopic transport coefficients into \(\lambda\), \(\kappa\), and \(\lambda_p\), leaving the remaining time dependence encoded by powers of the dimensionless variable \(w\). 
The parameter $\lambda$ measures the leading rotational stress source response relative to the susceptibility. The additional source correction is instead controlled by the product $\lambda\lambda_p$, while $\kappa$ controls the direct self feedback correction.

The two terms \(2\widetilde{\gamma}\theta\Gamma\xi_{\mu\nu}\) and
\(2\gamma_{\phi p\phi}\theta\xi_{\mu\nu}\) have the same tensorial structure,
as both are proportional to \(\theta\xi_{\mu\nu}\). 
We can also see from Eq.~\eqref{eq:master_attractor} that they affect the attractor equation in the same way up to the value of the corresponding coefficient combination $\lambda \lambda_p$. 
In the term by term attractor analysis, we therefore retain only one representative contribution of this type by using the benchmark normalization $\lambda_p=1$.

Although charge neutral conformal scaling fixes 
\(\kappa\equiv \widetilde{\gamma}_{\phi}/\tau_{\phi}=-1\) 
in the strict conformal limit~\cite{She:2025qri}, we keep \(\kappa\) explicit to trace the role of 
the \(\widetilde{\gamma}_{\phi}\tau_{\phi}/\tau^2\) correction in the early and 
late time attractor analysis, and one can also regard deviations from \(\kappa=-1\) as a 
phenomenological generalization. 
Such deviations probe the sensitivity of the attractor structure to second order nonlocal memory effects.

The susceptibility may be defined using \(\chi_s\), using \(\chi_\omega\) with respect to the dimensionful spin chemical potential \(\omega_{\mu\nu}\), or using \(\chi_\Omega\) with respect to the dimensionless spin potential \(\Omega_{\mu\nu}\). These three descriptions differ only by temperature dependent rescalings of the susceptibility, which are absorbed into the definition of \(\lambda\). Therefore, they lead to the same final attractor equation.

\subsection{Settings}
We now consider three cases, where Cases B and C isolate the effects of the two corrections separately.

\paragraph{Case A: minimal causal spin dynamics.}

The baseline case is obtained by setting $\lambda_p=0$ and $\kappa=0$.
Equation \eqref{eq:master_attractor} reduces to
\begin{align}
\frac{2}{3}\frac{f'}{w}
+\frac{f^2}{w^2}
+\frac{f}{w}
+
\left[
w^{-1}
-w^{-2}
+4\lambda
\right]
=0 .
\label{eq:attractor_A}
\end{align}

\paragraph{Case B: Spin force source correction.}

This case retains the source side correction proportional to $\theta\xi_{\mu\nu}$ while omitting the self feedback correction. Thus
\begin{equation}
\kappa=0 .
\end{equation}
The attractor equation becomes
\begin{align}
\frac{2}{3}\frac{f'}{w}
+\frac{f^2}{w^2}
+\frac{f}{w}
+
\left[
w^{-1}
-w^{-2}
+4\lambda 
+\frac{4\lambda}{3} w^{-1}
\right]
=0 .
\label{eq:attractor_B}
\end{align}

\paragraph{Case C: Dissipative self feedback correction.}

This case retains only the self feedback term proportional to
$\widetilde{\gamma}_{\phi}\theta\phi_{\mu\nu}$, corresponding to
\begin{equation}
\lambda_p=0.
\end{equation}
Using $\widetilde{\gamma}_{\phi}/(\tau w)=\kappa w^{-2}$, Eq.~\eqref{eq:master_attractor} gives
\begin{align}
\frac{2}{3}\frac{f'}{w}
+\frac{f^2}{w^2}
+\frac{f}{w}
-\kappa\frac{f}{w^{2}}
+
\left[
w^{-1}
-w^{-2}
+4\lambda
-\kappa w^{-2}
\right]
=0 .
\label{eq:attractor_C}
\end{align}
Since $\gamma$ is dissipative and $\chi_s$ is the spin susceptibility, we take
\begin{equation}
\gamma>0,\qquad
\chi_s>0,\qquad
\lambda>0 .
\end{equation}

\subsubsection*{Limiting cases for comparison}
The following ideal spin, Navier Stokes, and homogeneous relaxation limits are introduced for comparison and are not part of Cases A--C. They are used to distinguish the effects of relaxation from those of the second order corrections.
The ideal spin limit is obtained by switching off the rotational stress, $\phi=0$.
The spin conservation equation \eqref{eq:spin_reduced_1} then gives
\begin{equation}
\partial_\tau S+\frac{S}{\tau}=0,
\end{equation}
which implies
\begin{equation}
f_{\rm ideal}(w) = \frac{\tau}{S}\frac{dS}{d\tau} = -1.
\label{attractor_ideal}
\end{equation}

The Navier Stokes spin limit is obtained by neglecting the relaxation term
and all second order expansion corrections in the constitutive relation,
\begin{equation}
\partial_\tau\phi=0,
\qquad
\lambda_p=0,
\qquad
\kappa=0.
\end{equation}
The constitutive relation reduces to
\begin{equation}
\phi_{\rm NS}
=
2\gamma\frac{S}{\chi_s}.
\end{equation}
Substituting this constitutive relation into Eq.~\eqref{eq:spin_reduced_1} gives the Navier Stokes spin attractor function
\begin{equation}
f_{\rm NS}(w)
=
-1-4\lambda w .
\label{eq:attractor_E}
\end{equation}
This algebraic curve represents the first order dissipative spin limit.

We refer to the conservation laws coupled to the homogeneous part of the relaxation equations for dissipative currents as the homogeneous relaxation limit of second order viscous hydrodynamics.
It is obtained by removing the dissipative spin force source while keeping the second order propagation structure. 
We set
\begin{equation}
\lambda=0,
\qquad
\kappa=0 .
\end{equation}
Equation \eqref{eq:master_attractor} then becomes
\begin{align}
\frac{f^2}{w^2}
+\frac{2}{3}\frac{f'}{w}
+\frac{f}{w}
+
\left[
w^{-1}
-w^{-2}
\right]
=0 .
\label{eq:attractor_F}
\end{align}
Equivalently,
\begin{equation}
\frac{2}{3}w f'
+f^2
+w f
+w
-1
=0 .
\end{equation}
This equation contains the ideal Bjorken branch
\begin{equation}
f(w)=-1
\end{equation}
as an exact solution, while the full second order equation also allows
nontrivial transient evolution before approaching the ideal limit branch.

\section{Analysis of Attractor Behavior}
\label{section4}

\subsection{Early time fixed points and regularity selection}
 
We determine the early time fixed points by a dominant-balance analysis, in analogy with the small \(w\) fixed-point structure discussed in Ref.~\cite{Blaizot:2021cdv}.
We select the early time attractor candidate by requiring \(f(w)\) to remain regular near \(w=0\).
A branch is selected when its linear perturbation diverges as \(w\to0\), because regularity then fixes the perturbation amplitude to zero.
The case~A spin hydrodynamic equation Eq.~\eqref{eq:attractor_A} multiply $w^2$ gives the equivalent form
\begin{align}
    f^2 + \frac{2}{3} w f' + w f
    + \left[w - 1 + 4\lambda w^2\right]=0 .
    \label{eq:baseline_multiplied}
\end{align}

As $w\to 0$, the dominant balance of Eq.~\eqref{eq:baseline_multiplied} is
\begin{align}
    f_0^2 - 1 = 0 ,
\end{align}
which admits two early time fixed point candidates
\begin{align}
    f_0=\pm 1 .
\end{align}
We seek an asymptotic expansion
\begin{align}
    f(w) = f_0 + \sum_n a_n w^{\alpha_n},
\end{align}
and determine the allowed branches order by order.

\paragraph{Positive branch $(f_0=1)$.}
Substituting
\begin{align}
    f(w)=1+a_1 w+a_2 w^2+\mathcal{O}(w^3)
\end{align}
into Eq.~\eqref{eq:baseline_multiplied} gives
\begin{align}
    a_1=-\frac{3}{4},
    \qquad
    a_2=\frac{9}{160}-\frac{6}{5}\lambda .
\end{align}
Therefore the positive branch is
\begin{align}
    f_{+}(w)
    =
    1
    -\frac{3}{4}w
    +\left(
        \frac{9}{160}
        -\frac{6}{5}\lambda
    \right)w^2
    +\mathcal{O}(w^3).
    \label{eq:fplus_baseline}
\end{align}
Consider a linear perturbation around this branch,
\begin{align}
    f(w)=f_{+}(w)+\delta f(w).
\end{align}
At leading order as $w\to 0$, the perturbation satisfies
\begin{align}
    \frac{2}{3}w\,\delta f' + 2\,\delta f =0,
\end{align}
and hence
\begin{align}
    \delta f \sim w^{-3}.
\end{align}
This perturbation mode diverges as $w \to 0$ and is excluded by early time regularity. Thus the $f_0=1$ branch is an isolated regular early time solution.

\paragraph{Negative branch $(f_0=-1)$.}
Similarly, writing $f(w)=-1+\delta f(w)$, Eq.~\eqref{eq:baseline_multiplied} becomes
\begin{align}
    \frac{2}{3}w\,\delta f'
    -2\delta f
    +w\delta f
    +(\delta f)^2
    +4\lambda w^2
    =0 .
    \label{eq:baseline_negative_deltaf}
\end{align}
The linear operator
\begin{align}
    \mathcal{L}[\delta f]
    =
    \frac{2}{3}w\,\delta f'-2\delta f
\end{align}
has a homogeneous solution $\delta f\sim w^3$. Since the corrected forcing term is
$4\lambda w^2$, the leading particular solution is of order $w^2$.
Solving order by order gives
\begin{align}
    f_{-}(w)
    =
    -1
    +6\lambda w^2
    -9\lambda w^3\log w
    +Cw^3
    +\mathcal{O}\!\left(w^4\log^2 w\right),
    \label{eq:fminus_baseline}
\end{align}
where $C$ is an undetermined integration constant. The logarithm appears because
$w^3$ is the resonant power of the homogeneous mode.

For the negative branch, the perturbation behaves as $\delta f\sim w^3$, which remains finite as $w\to0$. Thus early time regularity does not exclude this perturbation, and the negative branch is not singled out as an isolated solution.
We therefore do not consider this branch further. 
This choice is also physically constrained by the Bjorken expansion setup, where the spin signal is assumed to start from a small initial value and is subsequently diluted by longitudinal expansion.

Finally, in Case A the isolated regular early time attractor candidate in the baseline case is
therefore
\begin{align}
    f_{\text{att}}^{(\text{A})}(w)
    =
    1
    -\frac{3}{4}w
    +\left(
        \frac{9}{160}
        -\frac{6}{5}\lambda
    \right)w^2
    +\mathcal{O}(w^3).
    \label{eq:fatt_base}
\end{align}

In case~B, including the driving source expansion term
$+\frac{4\lambda}{3w}$, multiplying Eq.~\eqref{eq:attractor_B} by $w^2$ gives
\begin{align}
    f^2
    + \frac{2}{3} w f'
    + w f
    + w -1
    +4\lambda w^2
    +\frac{4\lambda}{3}w
    =0 .
    \label{eq:source_driving_multiplied}
\end{align}
The additional source driving contribution is of order $w$ and is subleading compared with the constant balance $f_0^2-1$ as $w\to 0$.
Therefore the dominant early time balance remains
\begin{align}
    f_0^2-1=0,
\end{align}
with fixed-point candidates $f_0=\pm1$.

Therefore, the leading fixed-point structure and the regularity-based branch selection remain unchanged.
For the isolated positive branch, we write
\begin{align}
    f(w)
    =
    1
    +b_1 w
    +b_2 w^2
    +\mathcal{O}(w^3).
\end{align}
Solving Eq.~\eqref{eq:source_driving_multiplied} order by order gives
\begin{align}
    b_1=-\frac{\lambda}{2}-\frac{3}{4},
    \qquad
    b_2=
    \frac{9}{160}
    -\frac{51}{40}\lambda
    -\frac{3}{40}\lambda^2 .
\end{align}
Thus the source driving attractor candidate is
\begin{align}
    f_{\text{att}}^{(\text{B})}(w)
    =
    1
    -\left(
        \frac{\lambda}{2}+\frac{3}{4}
    \right)w
    +\left(
        \frac{9}{160}
        -\frac{51}{40}\lambda
        -\frac{3}{40}\lambda^2
    \right)w^2
    +\mathcal{O}(w^3).
    \label{eq:fatt_drive}
\end{align}
Hence the driving source expansion term modifies the subleading structure of the regular branch already at order $w$, but it does not change the dominant early time fixed point balance. For $\lambda>0$, it strengthens the downward departure of the attractor from the positive fixed point.

In case~C, the self feedback term $-\kappa(\frac{f}{w^{2}}+\frac{1}{w^{2}})$ is of the same order as the baseline $w^{-2}$ terms and therefore modifies the dominant algebraic fixed point
balance. Multiplying Eq.~\eqref{eq:attractor_C} by $w^2$ gives
\begin{align}
    \frac{2}{3} w f'
    + f^2
    -\kappa f
    -\kappa
    -1
    +w(f+1)
    +4\lambda w^2
    =0 .
    \label{eq:self_relax_multiplied}
\end{align}
If one writes $f(w)\to f_0$ as $w\to0$, the dominant early time balance is
\begin{align}
    f_0^2-\kappa f_0-\kappa-1=0,
\end{align}
or
\begin{align}
   (f_0+1)(f_0-\kappa-1)=0.   
\end{align}
Thus the finite early time fixed point candidates are
\begin{align}
    f_0=-1,
    \qquad
    \text{and}
    \qquad
    f_0=1+\kappa.
\end{align}
In the following power series expansions, we take generic negative values of \(\kappa\). This choice is motivated by the microscopic conformal limit result $\kappa=-1$ discussed in the previous section, while deviations from this value are treated there as a phenomenological generalization around the conformal limit value.

For the $f_0=1+\kappa$ branch, we write
\begin{align}
    f(w)
    =
    1+\kappa
    +c_1 w
    +c_2 w^2
    +\mathcal{O}(w^3).
\end{align}
Solving Eq.~\eqref{eq:self_relax_multiplied} order by order gives
\begin{align}
    c_1
    &=
    -\frac{6+3\kappa}{8+3\kappa},
    \\
    c_2
    &=
    -\frac{1}{\frac{10}{3}+\kappa}
    \left[
        -\frac{6(\kappa+2)}{(3\kappa+8)^2}
        +4\lambda
    \right].
\end{align}
Hence
\begin{align}
    f^{(\text{C})}(w)
    =
    1+\kappa
    -\frac{3(\kappa+2)}{3\kappa+8}w
    -\frac{1}{\frac{10}{3}+\kappa}
    \left[
        -\frac{6(\kappa+2)}{(3\kappa+8)^2}
        +4\lambda
    \right]w^2
    +\mathcal{O}(w^3).
    \label{eq:fatt_self}
\end{align}
A linear perturbation around this branch behaves as
\begin{align}
    \delta f
    \sim
    w^{-\frac{3}{2}(\kappa+2}) .
\end{align}
Since $-2<\kappa<0$, this perturbation diverges as \(w\to0\). Therefore the
$f_0=1+\kappa$ branch is isolated by early-time regularity.
Moreover, in the Bjorken expansion of heavy ion collisions, the early time spin signal is expected to be generated from a small initial value before being diluted by the subsequent longitudinal expansion. 
A positive $f_0$ occurs on this branch only for $\kappa>-1$. This branch approaches $f_0=0$ in the strict conformal limit $\kappa=-1$.

\begin{table*}[h]
\centering
\renewcommand{\arraystretch}{1.35}
\setlength{\tabcolsep}{2pt}
\footnotesize
\begin{tabular}{|c|c|c|c|c|}
\hline
Case
& \shortstack[c]{Fixed point(s)}
& \shortstack[c]{Selected branch}
& \shortstack[c]{Leading\\correction}
\\
\hline

Ideal limit
& $f_0=-1$ 
& $f_0=-1$
&  N/A
\\
\hline

Navier--Stokes limit
& $f_0=-1$
& $f_0=-1$
& $-4\lambda w$
\\
\hline

homogeneous relaxation
& \shortstack[l]{$f_0=\pm 1$}
& $f_0=1$
& $-\frac{3}{4}w$
\\
\hline

A: baseline
& $f_0=\pm 1$
& $f_0=1$
& $-\frac{3}{4}w$
\\
\hline

B: source driving
& $f_0=\pm 1$
& $f_0=1$
& $-\left(\frac{3}{4}+\frac{\lambda}{2}\right)w$
\\
\hline

C: self feedback
& \shortstack[l]{$f_0=-1$,\\$f_0=1+\kappa$}
& \shortstack[l]{$f_0=1+\kappa$}
& \shortstack[l]{$-\frac{3(\kappa+2)}{3\kappa+8}w$}
\\
\hline
\end{tabular}
\caption{
Early time fixed structures, selected branches and leading correction for different cases.
}
\label{tab:early_time_summary_all_cases}
\end{table*}

We summarize the early time structure of the spin attractor in
Table~\ref{tab:early_time_summary_all_cases}, including the ideal
spin hydrodynamic limit, the Navier Stokes limit, the homogeneous relaxation
limit, and the three dynamical cases considered in this work. 
The table separates three pieces of early time information: the fixed point
structure, the selected attractor branch, and the leading correction away from
that branch.

The homogeneous second order relaxation structure is responsible for the
appearance of the two finite early time fixed points $f_0=\pm1$. 
This double fixed point structure is absent in the ideal and Navier Stokes
limits, where the spin evolution is controlled by the first order
dilution/relaxation balance and the only fixed point is $f_0=-1$. 
Once the second order propagation term is retained, however, the early time
equation admits both $f_0=1$ and $f_0=-1$. 
In the spin probe regime, we focus on the positive branch $f_0=1$, for which
$f=(2/3)d\ln S/d\ln w>0$ and therefore $S(w)\sim w^{3/2}$ as $w\to0$.
When the nonlocal memory induced self feedback correction is included, however, this fixed point is shifted from $f_0=1$ to $f_0=1+\kappa$, and the corresponding early time scaling also changes.

Cases A and B are perturbative deformations of this homogeneous second order
relaxation structure. 
Case A preserves both the fixed point structure and the leading departure from
the selected positive branch,
\begin{equation}
    f_{\rm att}^{({\rm A})}(w)
    =
    1-\frac{3}{4}w+\mathcal{O}(w^2).
\end{equation}
Thus the baseline dissipative correction proportional to $\lambda$ does not
affect the fixed point position or the first correction; it enters only at the
next order in the early time expansion.

Case B shows a different effect. 
The spin force source term is still subleading with respect to the dominant
fixed point balance, so it does not move the fixed points $f_0=\pm1$ and does
not change the selected leading scaling $f\sim1$. 
However, it contributes at the same order as the first finite $w$ correction. 
As a result, the leading departure becomes
\begin{equation}
    f_{\rm att}^{({\rm B})}(w)
    =
    1-\left(
        \frac{\lambda}{2}+\frac{3}{4}
    \right)w+\mathcal{O}(w^2).
\end{equation}
Therefore the source driving correction changes the magnitude of the initial
departure from the fixed point and enhances the downward bending of the
attractor for $\lambda>0$, but it does not create a new early time scaling
regime.

Case C is qualitatively different, but in a different way from Cases A and B. 
In case~C, the self feedback term is of the same order as the
baseline $w^{-2}$ terms.  
It therefore modifies the dominant algebraic fixed point balance rather than
generating a divergent early time branch. 
The leading balance becomes
\begin{equation}
    f_0^2-\kappa f_0-\kappa-1=0.
\end{equation}
Thus the case~C fixed point candidates are
\begin{equation}
    f_0=-1,
    \qquad
    f_0=\kappa + 1.
\end{equation}
The $f_0=1+\kappa$ branch remains finite and has the expansion
\begin{equation}
    f_{\rm att}^{({\rm C})}(w)
    =
    1+\kappa
    -\frac{3(\kappa+2)}{3\kappa+8}w
    +\mathcal{O}(w^2).
\end{equation}
Thus the self feedback correction reorganizes the finite fixed-point structure
and modifies the leading departure from the selected $f_0=1+\kappa$ branch, but it
does not produce the singular scaling $f\sim\kappa w^{-1/2}$ in the corrected
case~C equation.

The role of $\lambda$ also depends on the sector under consideration. 
In the Navier--Stokes limit, $\lambda$ directly controls the leading
dissipative correction,
\begin{equation}
    f_{\rm NS}(w)=-1-4\lambda w .
\end{equation}
In case~B, $\lambda$ enters the leading correction to the positive second order
branch through the source driving term and strengthens the downward departure
from the fixed point. 
In case~C, by contrast, $\lambda$ does not change the fixed point positions and
does not enter the first correction around the selected $f_0=1+\kappa$ branch. 
It first contributes at the next order in the finite early time expansion. 
This hierarchy shows that source driving modifies the first departure from the
finite fixed point, whereas the self feedback term, through $\kappa$, changes
the algebraic fixed point structure itself.

The early-time attractor analysis restricts the physical parameter regime to $\lambda > 0$, since only positive $\lambda$ leads to dissipative suppression of the spin density in the Navier–Stokes limit.

Microscopically, the early time behavior reflects the different roles of nonlinear response and nonlocal memory effects. The nonlinear response contribution enters through the source correction and mainly modifies the evolution away from the fixed point, whereas the nonlocal memory induced self feedback correction can shift the fixed point itself and thereby change the associated early-time scaling.

\subsection{Late-time decay and asymptotic behavior}

We now analyze the late time behavior of the full spin attractor equation Eq.(\ref{eq:master_attractor}),
\begin{align}
    \frac{f^2}{w^2}
    + \frac{2}{3}\frac{f'}{w}
    + \frac{f}{w}
    - \frac{\kappa f}{w^{2}}
    +\left[
        w^{-1}
        - w^{-2}
        +4\lambda
        +\frac{4\lambda}{3w}
        -\frac{\kappa}{w^{2}}
    \right]
    =0 .
    \label{eq:late_full_corrected}
\end{align}
In the notation of Ref.~\cite{Wang:2024afv}, the authors parametrize the conformal scaling variable and the strength of the rotational source as
\[
w=(\tau/\tau_1)^{\Delta_1},
\qquad
\tau_\phi\gamma/\chi=\alpha w^{\Delta_2}.
\]
Thus $\Delta_1$ is the power governing the proper time dependence of the inverse Knudsen variable $w$, while $\Delta_2$ is the power governing the $w$ dependence of the source-response combination $\tau_\phi\gamma/\chi$. In our conformal setup $w\propto\tau^{2/3}$ and $\lambda=\tau_\phi\gamma/\chi_\omega$ is constant, so $\Delta_1=2/3$ and $\Delta_2=0$. Comparing the constant source terms gives $8\alpha=4\lambda$, i.e. $\alpha=\lambda/2$.

In the present notation, after multiplying
Eq.~\eqref{eq:late_full_corrected} by $w^2$ and omitting the source driving
and self feedback terms, the equation reduces to the dominant balance equation of case~A. 
It therefore belongs to the same
\(\Delta_2=0\) late time dominant-balance class discussed in
Ref.~\cite{Wang:2024afv}, with the identifications \(\Delta_1=2/3\) and
\(\alpha=\lambda/2\). 
We may therefore take
\begin{align}
    f_{\pm}(w)\sim a_{\pm} w,
    \qquad
    a_{\pm}=\frac{-1\pm\sqrt{1-16\lambda}}{2}.
    \label{eq:late_a_roots}
\end{align}
and focus on how the source driving term from case~B and the self feedback terms from case~C modify late time behaviors.
Furthermore, we find the source driving and self feedback terms do not alter the dominant late time balance, so the leading branches remain \(f\sim a_\pm w\). 
The leading late time asymptotic structure has already been analyzed for the minimal causal spin hydrodynamic equation in Ref.~\cite{Wang:2024afv} .

To extract the first subleading correction from source driving term from case~B and the self feedback terms from case~C, write
\begin{align}
    f(w)
    \sim
    a w
    + b .
\end{align} 
At order $w^{-1}$,the minimal causal spin hydrodynamic equation, corresponding to Case A, contribution is
\begin{align}
    \left[
        (2a+1)b+\frac{2a}{3}+1
    \right]w^{-1},
\end{align}
where the derivative term gives
\begin{align}
    \frac{2}{3}\frac{f'}{w}
    \sim
    \frac{2a}{3}w^{-1},
\end{align}
and the correction $b$ contributes through
\begin{align}
    \frac{f^2}{w^2}+\frac{f}{w}
    =
    a^2+a
    +(2a+1)b\,w^{-1}
    +\cdots .
\end{align}

The case~B source-driving correction contributes
\begin{align}
    \frac{4\lambda}{3w}
    =
    \frac{4\lambda}{3}w^{-1}.
\end{align}
The case~C self feedback term contributes at the same order,
\begin{align}
    -\frac{\kappa f}{w^2}
    \sim
    -\kappa a\,w^{-1}.
\end{align}
The remaining case~C term $-\kappa/w^2$ contributes only at order $w^{-2}$.
Therefore, the coefficient \(b\) is fixed by the full \(w^{-1}\) balance, receiving contributions from the derivative term, the source driving correction, and the self feedback term.

The coefficient equation is
\begin{align}
    (2a+1)b
    +\left(\frac{2}{3}-\kappa\right)a
    +1
    +\frac{4\lambda}{3}
    =0 .
\end{align}
Using Eq.~\eqref{eq:late_a_roots}, this gives
\begin{align}
    b_{\pm}
    =
    \frac{\kappa}{2}
    -\frac{1}{3}
    \mp
    \frac{8\lambda+4+3\kappa}
    {6\sqrt{1-16\lambda}} .
\end{align}
For the ansatz $f\sim aw+b$, the case $\lambda=1/16$ is excluded because the two branches at leading order merge and parameter $b$ is not fixed at the first subleading order.

Consequently, for $\lambda\neq 1/16$, the late-time branches are
\begin{align}
    f_{\pm}(w)
    =
    \frac{-1\pm\sqrt{1-16\lambda}}{2}\,w
    +
    \left[
        \frac{\kappa}{2}
        -\frac{1}{3}
        \mp
        \frac{8\lambda+4+3\kappa}
        {6\sqrt{1-16\lambda}}
    \right]
    +\mathcal{O}\!\left(w^{-1}\right).
    \label{eq:late_f_generic}
\end{align}
Thus the corrected source driving term $4\lambda/(3w)$ and the
self feedback terms $-\kappa f/w^2-\kappa/w^2$ enter the subleading late time slopes $b_\pm$. 
Among these corrections, $4\lambda/(3w)$ and $-\kappa f/w^2$ enter the first subleading coefficients $b_\pm$, whereas $-\kappa/w^2$ contributes only at higher order.

From Eq.(\ref{eq:f_definition}), the amplitude $S(w)$ is related to $f(w)$ by
\begin{align}
    S(w)\sim
    \exp\left[
        \frac32\int^w \frac{f(\tilde w)}{\tilde w}\,d\tilde w
    \right].
\end{align}
Using Eq.~\eqref{eq:late_f_generic}, we obtain
\begin{align}
    \frac{f_{\pm}(w)}{w}
    =
    a_{\pm}
    +b_{\pm}w^{-1}
    +\mathcal{O}\!\left(w^{-2}\right),
\end{align}
and therefore
\begin{align}
    S_{\pm}(w)
    \sim
    \exp\left[
        \frac{3a_{\pm}}{2}w
        +\frac{3b_{\pm}}{2}\log w
        +\mathcal{O}\!\left(w^{-1}\right)
    \right].
    \label{eq:late_S_generic}
\end{align}

\begin{figure}[h]
    \includegraphics[width=0.6\linewidth]{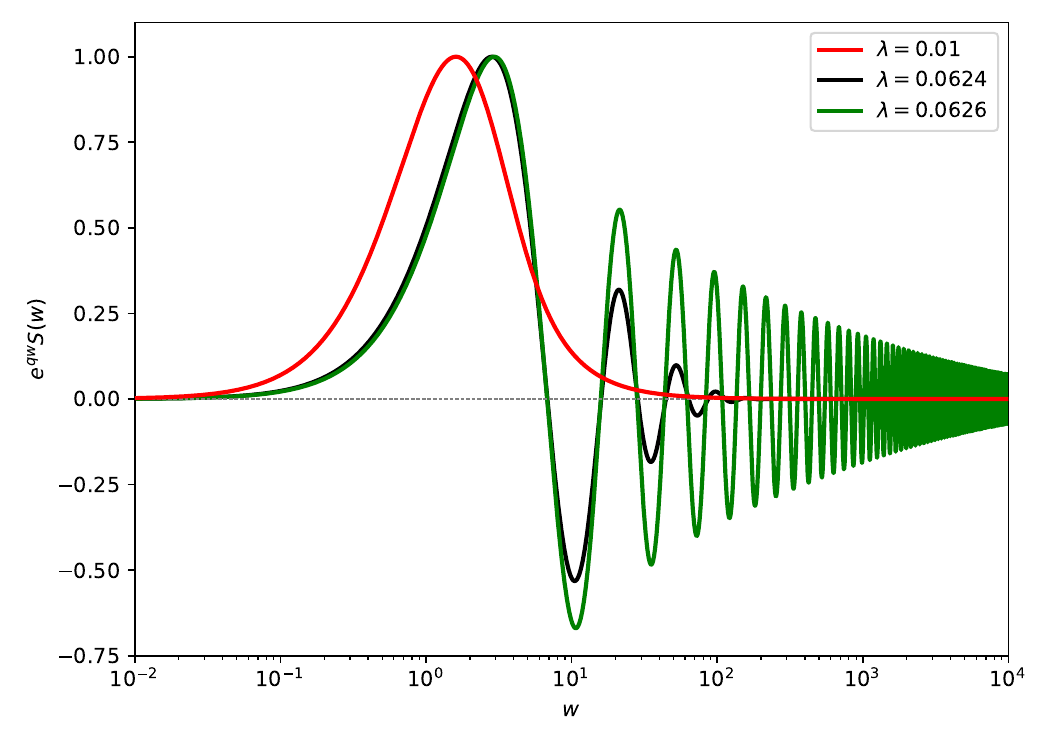}
    \caption{Numerical evolution of the rescaled spin density $e^{qw}S(w)$ in Case A for representative values of $\lambda$ below, close to, and above the critical value $\lambda=1/16$. For $\lambda>1/16$, the late time solution becomes oscillatory after removing the leading exponential damping, reflecting the exponentially damped oscillation of $S(w)$.}
    \label{fig:S_late_time}
\end{figure}

For positive $\lambda$, the qualitative late time behavior is controlled by the sign of the discriminant $(1-16\lambda)$.  We consider two regimes $\lambda <1/16$ and $\lambda > 1/16$.
To further illustrate the late time behavior on the two sides of the critical value
$\lambda=1/16$, we numerically solve the Case A equation for several representative
values of $\lambda$. In order to make the late time behavior more visible, we plot the
rescaled spin density $e^{qw}S(w)$, where
$q=\frac{3}{4}(1-\sqrt{1-16\lambda})$ for $\lambda<1/16$ and
$q=3/4$ for $\lambda>1/16$.
Figure~\ref{fig:S_late_time} shows the numerical results for
$\lambda=0.01$, $0.0624$, and $0.0626$.
For $\lambda=0.01$, the solution approaches the late time regime without oscillation.
For $\lambda=0.0624$, which is close to the critical value, the solution shows a stronger
transient behavior before approaching the late time regime.
For $\lambda=0.0626>1/16$, clear oscillations appear at late times.
This agrees with the analytical result that the late time solution is non-oscillatory
for $\lambda<1/16$, while it becomes an exponentially damped oscillatory mode for
$\lambda>1/16$.

First, for $\lambda <1/16$, the two possible leading coefficients $a_{\pm}$ are real.
Using Eq.~\eqref{eq:late_S_generic}, the corresponding amplitudes are
\begin{align}
    S_{\pm}(w)
    &\sim
    \exp\left[
        \frac{3a_{\pm}}{2}w
        +\frac{3b_{\pm}}{2}\log w
        +\mathcal{O}\!\left(w^{-1}\right)
    \right] \nonumber\\
    &=
    w^{\frac{3b_{\pm}}{2}}
    \exp\left[
        \frac{3a_{\pm}}{2}w
        +\mathcal{O}\!\left(w^{-1}\right)
    \right].
\end{align}
Both branches are exponentially damped. Since $a_{+}>a_{-}$, the $a_{+}$ branch is the less damped one and therefore dominates at sufficiently late times. 
In this regime, the late time behavior is non-oscillatory. 
The corrected source driving and self feedback terms do not change the exponential damping rates, but they enter the coefficients $b_{\pm}$ and therefore generate the power law prefactor.

Second, for $\lambda > 1/16$, the two slopes $a_{\pm}$ form a complex conjugate pair. 
Defining
\begin{align}
    \nu=\sqrt{16\lambda-1},
\end{align}
Eq.~\eqref{eq:late_f_generic} becomes
\begin{align}
    f_{\pm}(w)
    =
    -\frac12 w
    \pm \frac{i\nu}{2}w
    +\frac{\kappa}{2}
    -\frac{1}{3}
    \pm i\,\frac{8\lambda+4+3\kappa}{6\nu}
    +\mathcal{O}\!\left(w^{-1}\right).
    \label{eq:late_f_oscillatory}
\end{align}
The corresponding amplitudes behave as
\begin{align}
    S_{\pm}(w)
    &\sim
    \exp\left[
        -\frac{3}{4}w
        \pm i\frac{3\nu}{4}w
        +\left(
            \frac{3\kappa}{4}
            -\frac12
        \right)\log w
        \pm i\frac{8\lambda+4+3\kappa}{4\nu}\log w
        +\mathcal{O}\!\left(w^{-1}\right)
    \right] \nonumber\\
    &=
    w^{\frac{3\kappa}{4}-\frac12}
    e^{-3w/4}
    \exp\left[
        \pm i\frac{3\nu}{4}w
        \pm i\frac{8\lambda+4+3\kappa}{4\nu}\log w
        +\mathcal{O}\!\left(w^{-1}\right)
    \right].
    \label{eq:late_S_oscillatory}
\end{align}
Thus, for \(\lambda>1/16\), the late time behavior is an exponentially damped
oscillatory mode. 
The leading attenuation factor is \(e^{-3w/4}\), which is unchanged by the source and feedback corrections. 
The oscillation frequency $\nu$ is determined by the leading order balance of the baseline Case A equation rather than by the additional source correction. The source correction affects the solution only at subleading order, through the constant term in $f$ and the corresponding logarithmic phase shift of $S$. In contrast, the feedback correction modifies the algebraic prefactor \(w^{3\kappa/4-1/2}\) and also contributes to the logarithmic phase shift.

We conclude that the additional source driving and self feedback terms do not modify the leading late time dominant balance of the minimal causal spin hydrodynamic equation~\cite{Wang:2024afv}.
After the identifications $\Delta_1=2/3$ and $\alpha=\lambda/2$,  the leading branches are governed by the $\Delta_2=0$ balance of Ref.~\cite{Wang:2024afv}, and hence the leading coefficients $a_\pm$ are unchanged.

These terms, however, modify the late time solution at the first subleading order. 
This is manifest in Eq.~\eqref{eq:late_f_generic}, where the constant coefficients $b_\pm$ receive explicit corrections from the source driving and self feedback contributions. 
As a result, the amplitude acquires an algebraic prefactor. 
For \(0<\lambda<1/16\), the two branches have real leading coefficients and the late time behavior is non-oscillatory, with the least damped branch controlled by \(a_+\). 
For \(\lambda>1/16\), the two branches form a complex conjugate pair, leading to exponentially damped oscillations. 
In this regime, the source correction contributes to the oscillatory phase, while the feedback
correction modifies the algebraic prefactor; both corrections enter the logarithmic phase shift.
For the ansatz $f\sim aw+b$, the case $\lambda=1/16$ is excluded because the two branches at leading order merge and parameter $b$ is not fixed at the first subleading order.

In microscopic terms, nonlinear response effects enter through the source correction and therefore affect the late time solution only at subleading order, while nonlocal memory effects enter through both the source and self feedback corrections and modify the subleading phase and algebraic prefactor. Neither effect changes the leading late-time damping structure.

\subsection{Numerical Evolution and Attractor Convergence}

\begin{figure}[h]
    \includegraphics[width=0.4729\linewidth]{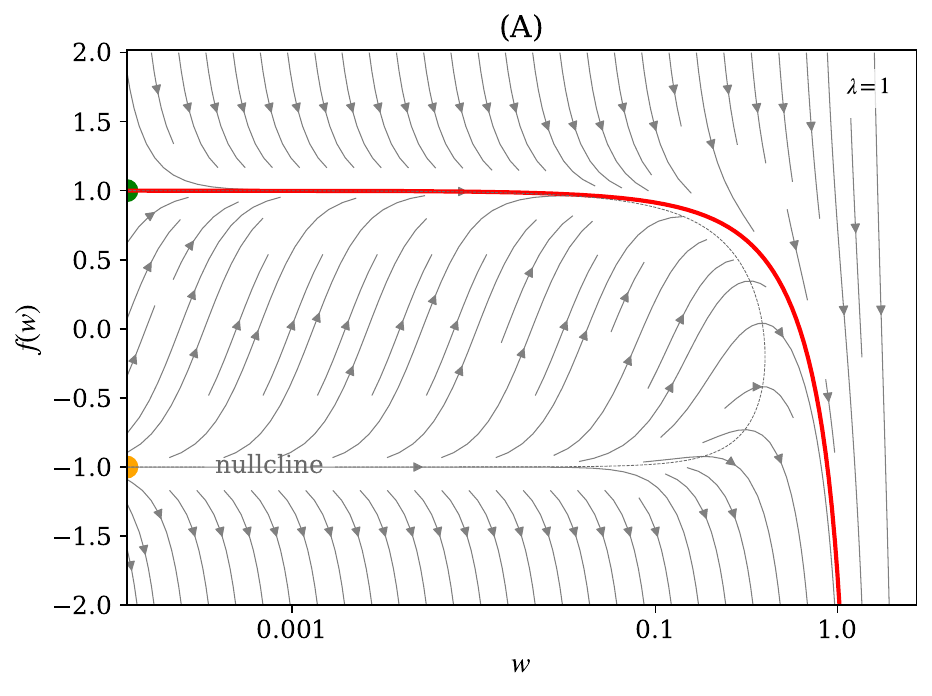}
    \includegraphics[width=0.4729\linewidth]{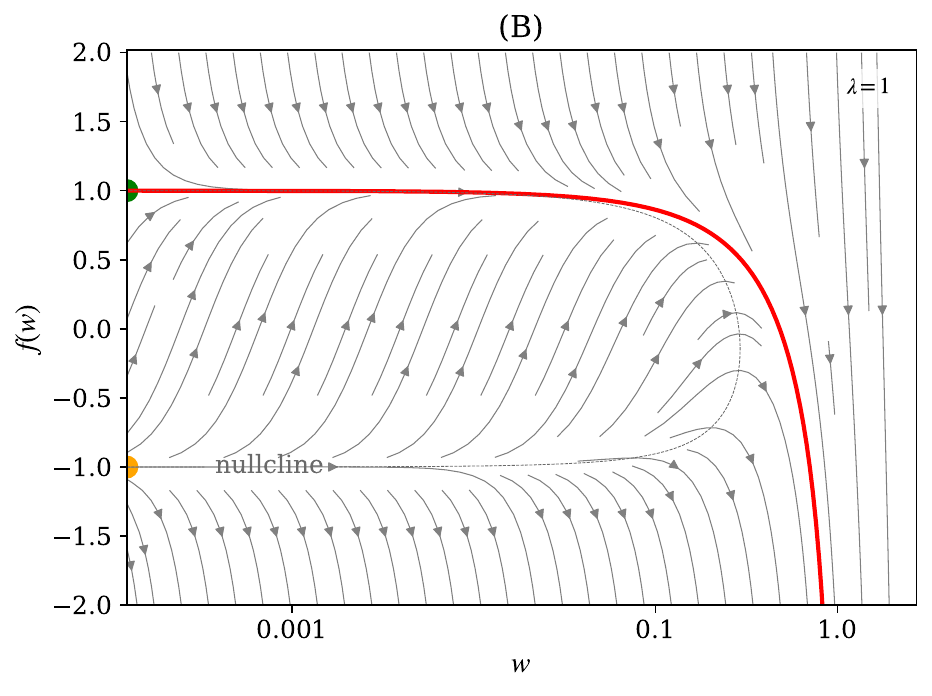}
    \includegraphics[width=0.4729\linewidth]{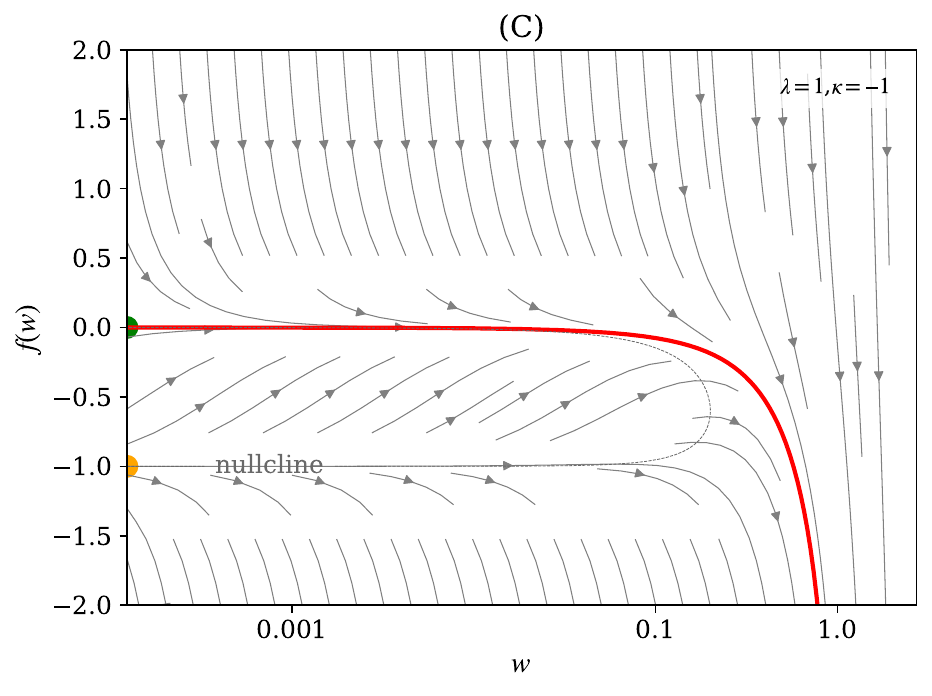}
    \includegraphics[width=0.4729\linewidth]{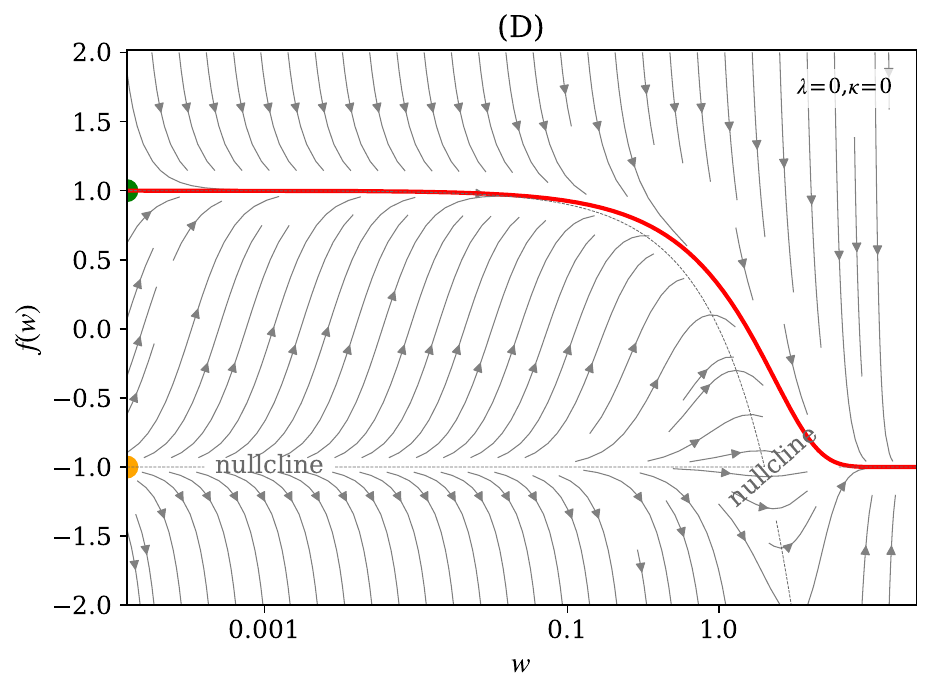}
    \caption{Attractor solutions for Case A, B, and C.
    Arrows show the flow fields, red solid curves denote attractors, and filled circles mark fixed points. Cases A and B depend on \(\lambda\), whereas Case C depends on \(\lambda\) and \(\kappa\).}
    \label{fig:fixpoint_streamlines}
\end{figure}

In the conformal limit, we take the spin relaxation time to be  \(\tau_\phi(\tau)=c_\phi T^{-1}(\tau)\). 
Since the attractor equations are controlled by the scaling structure, the overall constant \(c_\phi\) can be absorbed into the choice of conformal units. 
It is then convenient to formulate the numerical analysis in terms of dimensionless transport combinations. 
The parameter $\lambda$ is the baseline dimensionless source response combination $\tau_\phi\gamma/\chi_\omega$ and is present already in Case A. The additional source correction in Case B is controlled by $\lambda\lambda_p$; after adopting the benchmark normalization $\lambda_p=1$, its coefficient is numerically proportional to $\lambda$. The parameter $\kappa$ denotes the dimensionless coefficient associated with the self feedback correction. 

Figure~\ref{fig:fixpoint_streamlines} presents the phase space structure of the three spin cases together with the homogeneous relaxation limit. Panels (a)--(c) correspond to Cases A, B, and C, respectively, whereas panel (d) shows the homogeneous relaxation limit with \(\lambda=0\) and \(\kappa=0\). 
In Cases A and B, the early-time fixed points are located at \(f=1\) and \(f=-1\).
For Case C, however, the fixed points are \(f=-1\) and \(f=1+\kappa\). 
This shift shows that the feedback correction qualitatively changes the early fixed point structure of the attractor equation in Case C. 
It further suggests that the second order correction in NESO spin hydrodynamics can significantly affect the early time transient dynamics by shifting the $f=1$ fixed point to $f=1+\kappa$ while leaving the $f=-1$ fixed point unchanged.
In contrast, the correction of the source in Case B does not lead to a comparable modification of the early time dynamics.

As discussed in the previous subsection, the large \(w\) behavior of \(f(w)\) is characterized by exponential damping rather than by a stable late time hydrodynamic fixed point. 
This is in contrast to the homogeneous relaxation limit of spin hydrodynamics in plot(D) of Figure \ref{fig:fixpoint_streamlines} and to the ideal spin hydrodynamic limit in
Eq.~\eqref{attractor_ideal}, where the late time hydrodynamic fixed point remains robust in the conformal limit. 
This comparison indicates that the source terms on the right hand side of the relaxation equation Eq.(\ref{relaxation_phi}) qualitatively modify the large \(w\) tail and can destroy the stable asymptotic behavior present in the source free limits.

\begin{figure}[h]
    \includegraphics[width=0.6\linewidth]{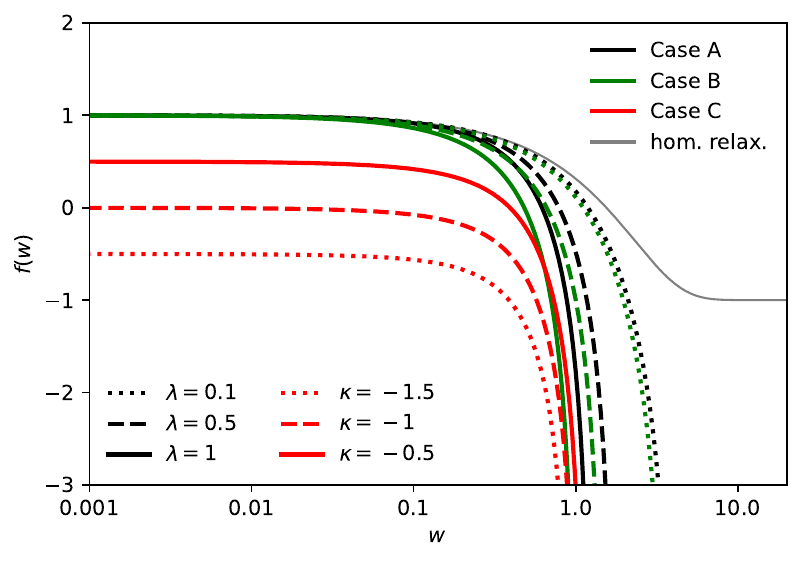}
    \caption{Comparison of the numerical attractor solutions for Cases A, B, and C with the homogeneous relaxation limit. In Cases A and B, different values of $\lambda$ modify the evolution away from the early time fixed point $f_0=1$. In Case C, varying $\kappa$ shifts the early time fixed point according to $f_0=1+\kappa$.}
    \label{fig:comparison_ABC}
\end{figure}

Figure~\ref{fig:comparison_ABC} compares the numerical solutions of Cases A, B, and C with the homogeneous relaxation limit. In Cases A and B, varying $\lambda$ changes the subsequent evolution of the attractor solution, while the early time fixed point remains at $f_0=1$. The source correction in Case B leads to a stronger departure from the early time fixed point compared with Case A. In Case C, the early time behavior depends directly on $\kappa$, since the corresponding fixed point is $f_0=1+\kappa$. For the representative values $\kappa=-1.5$, $-1$, and $-0.5$, the fixed point therefore shifts to $f_0=-0.5$, $0$, and $0.5$, respectively. This numerical comparison illustrates that the self feedback correction changes the early time fixed point structure, whereas the corrections in Cases A and B mainly modify the evolution away from the fixed point. 

In terms of their microscopic origins, nonlinear response effects mainly modify the evolution away from the early-time fixed point through source corrections, whereas nonlocal memory effects can affect both the subsequent evolution and, through the self feedback correction, the early time fixed point structure itself.

\section{Summary}
\label{section5}
In this work, we studied attractor dynamics in spin hydrodynamics in the spin probe regime within Zubarev's non-equilibrium statistical operator formalism. 
Specializing the theory to \((0+1)\)D Bjorken flow, we derived the corresponding attractor equations and examined how rotational and dissipative corrections modify the evolution of the spin sector.
The dissipative spin flux contribution vanishes, leaving local spin orbital exchange mediated by the antisymmetric rotational stress tensor as the relevant relaxation channel.

A central reference point of our analysis is the minimal causal spin hydrodynamic attractor, referred to here as case~A. While this framework has been discussed previously, its early time fixed point structure has not been analyzed in detail. 
We showed that the underlying second order relaxation structure admits two finite early time fixed points and that the positive branch provides the relevant early time attractor solution in the spin probe limit.

We then considered corrections to this baseline attractor structure. 
In case~B, the source driving term preserves the fixed point pattern of case~A but modifies the leading correction to attractor solution, thereby changing the initial departure from the selected branch. 
In case~C, by contrast, the self feedback term enters the dominant algebraic balance itself and reorganizes the early time fixed point structure. 
These results show that different spin corrections affect the attractor in qualitatively different ways: source driving deforms the approach to a given branch, whereas self feedback can alter the fixed point structure itself.
At late times, the leading asymptotic branches remain those of the minimal causal spin hydrodynamic attractor and are therefore not affected by the source driving and self feedback terms. 
The new effect of these terms appears at the first subleading order. 
Thus, while the dominant late time damping structure is unchanged, the subleading asymptotic behavior retains clear sensitivity to the additional closure dependent corrections.

Overall, our results provide a systematic extension of the minimal causal spin hydrodynamic attractor framework. 
We identify the early time fixed point structure of the baseline theory, clarify how source driving and self feedback modify the attractor under Bjorken expansion, and show how these terms affect the subleading late time asymptotics without changing the leading late time branches. In microscopic terms, nonlinear response effects contribute to the source correction that modifies the evolution away from the early time fixed point, whereas nonlocal memory effects contribute to both the source and self feedback corrections and can therefore also modify the fixed point structure itself.

These conclusions are obtained under the assumptions of the spin probe approximation, charge neutrality, an ideal conformal Bjorken background, and the adopted pseudogauge. The loss of initial condition memory through relaxation and the regularity based selection of the attractor branch are expected to be relatively robust, since they are mainly controlled by the dissipative relaxation structure. 
In contrast, the precise locations of fixed points, subleading coefficients, and the relative roles of nonlinear response and nonlocal memory effects may depend more sensitively on the hydrodynamic closure, background evolution, and pseudogauge choice. 
Further extensions including spin backreaction, more general hydrodynamic backgrounds, and alternative second order closures and pseudogauges would therefore be useful for testing the robustness of the attractor structure. In particular, such comparisons can help distinguish the leading late-time damping and regularity selected branch from more model dependent features such as shifted fixed points and logarithmic phase corrections.

\section*{ACKNOWLEDGMENTS}
Q. Zhou would like to thank J. Zhu for helpful discussions.
This research is supported by the National Natural Science Foundation of China with Project Nos.~12635009, Nos.~12605222 and Nos.~12535010.
Duan She's research is funded by the Startup Research Fund of Henan Academy of Sciences (No. 231820058), the High level Achievements Reward and Cultivation Projects (No. 20252320001), the Basic Research Fund of Henan Academy of Sciences (No.20260620005), and the Key Laboratory of Quark and Lepton Physics (No. QLPL2025P01).

\bibliography{refs}
\end{document}